\documentclass[aps,reprint,amsmath,amssymb,longbibliography,floatfix]{revtex4-2}

\usepackage[T1]{fontenc}
\usepackage{lmodern}
\usepackage[utf8]{inputenc}
\usepackage{amsmath,amssymb,amsthm,bm,mathtools}
\usepackage{graphicx}
\usepackage{booktabs}
\usepackage{array}
\usepackage{xcolor}
\definecolor{UpdateGreen}{RGB}{0,112,48}
\newcommand{\update}[1]{{\color{UpdateGreen}#1}}

\graphicspath{{figures/}}
\usepackage{slashed}
\usepackage[hypertexnames=false,colorlinks=true,linkcolor=black,citecolor=black,urlcolor=black]{hyperref}

\newcommand{\nn}{\nonumber}
\newcommand{\bea}{\begin{eqnarray}}
\newcommand{\ena}{\end{eqnarray}}

\newcommand{\ba}{\begin{array}}
\newcommand{\ea}{\end{array}}
\newcommand{\be}{\begin{equation}}
\newcommand{\ee}{\end{equation}}

\begin{document}
\title{Critical properties of the two-dimensional fully anisotropic Ising model on various lattices}

\author{Sh.~Khachatryan}
\author{A.~Sedrakyan}
\affiliation{Alikhanyan National Science Laboratory, Br.\ Alikhanian 2, Yerevan 0036, Armenia}

\begin{abstract}
We develop a unified fermionic-field formulation of the
	two-dimensional Ising model on several planar lattices using the
	Kac--Ward representation. Grassmann fields are associated with directed
	lattice links, while changes in the direction of fermionic trajectories
	are encoded by the corresponding Kac--Ward phase factors. The resulting
	quadratic action reduces, after Fourier transformation, to a
	finite-dimensional momentum-space fermionic matrix. Its determinant defines the fermionic spectral polynomial, while its zero modes determine the critical surface.
	We apply this construction to the square, honeycomb, triangular,
	kagom\'e, and dual kagom\'e (dice or $T_3$) lattices. The known critical
	couplings of the homogeneous models are reproduced as consistency
	checks. For the fully anisotropic kagom\'e model with three independent
	couplings we recover the previously known fully anisotropic {kagom\'e} critical surface within the present Kac–Ward construction and the corresponding
	low-energy spectral equation. We construct the analogous $12\times12$
	fermionic matrix for the anisotropic dice lattice and derive its critical
	surface, which is related to that of the kagom\'e lattice by the Ising
	duality transformation.
The same fermionic formulation yields the physical spin correlations. For
kagom\'e, the full correlation Pfaffian separates into a common bulk
Toeplitz factor and a local sublattice amplitude,
$M_i=\rho_i(1-\kappa_K^2)^{1/8}$, with $\rho_i$ expressed directly in
$\lambda_i=\tanh J_i$. The independent star--triangle check gives
$\rho_i=\tanh(2L_i)$ and reproduces Naya's isotropic result.
For dice, the same reduction gives $M_1=(1-\kappa_D^2)^{1/8}$ for the
six-coordinated sublattice and
$M_2=M_3=A_D M_1+B_D T_\triangle$ for the two three-coordinated
sublattices, where $T_\triangle$ is the elementary triplet order parameter
of the reduced triangular model. Isotropically,
$M_2=M_3=\rho_3^D M_1$.
Thus the same fermionic construction provides a common framework for
	calculating critical surfaces, excitation spectra, Green functions,
	spin correlations, and spontaneous magnetization for Ising models on
	planar lattices with different geometries and coordination numbers.
\end{abstract}


\maketitle

\section{Introduction}

The two-dimensional Ising model (2DIM) is one of the fundamental
exactly solvable models of statistical mechanics. Originally introduced
as a simple model of ferromagnetism, it became a basic theoretical
laboratory for the study of phase transitions, critical phenomena,
universality, lattice statistical mechanics, and quantum field theory.
The exact solution of the square-lattice model obtained by Onsager
\cite{Onsager-1944}, together with Kaufman's spinor formulation
\cite{Kaufman-2}, provided the first exact description of a
finite-temperature phase transition in a system with short-range
interactions and established many of the mathematical structures which
later became central in the theory of exactly solvable and integrable
models.

An important step preceding Onsager's solution was the duality
transformation introduced by Kramers and Wannier
\cite{Kramers-Wannier-1941}. It relates the high- and low-temperature
regions of the Ising model and determines the critical point of the
square lattice from its self-duality. The same general idea of relating
different planar lattices by duality, decoration, and star--triangle
transformations subsequently became one of the main tools for the exact
solution of Ising models on more complicated two-dimensional graphs
\cite{Houtappel-1950,Syozi-1951,Kano-1953,Naya-1954,Syozi-1955,
Syozi-1960,Syozi-1960-2,Syozi-1972,Baxter-book,Chan,Perk,Au-Yang}. In particular, exact
results were obtained for the honeycomb and triangular lattices and
later for a broad class of decorated and inhomogeneous lattices.

A second, conceptually different, route to the exact solution is based
on the high-temperature graphical expansion. In this representation,
the Ising partition function is expressed as a sum over closed polygonal
configurations. Kac and Ward showed that the over-counting associated
with intersections of closed trajectories can be removed by assigning
a phase determined by the total rotation of the tangent vector
\cite{Kac-Ward-1952}. As a result, the partition function can be
represented in terms of the determinant of a matrix acting in the
space of directed lattice links. This construction provides a
particularly transparent connection between the geometry of the
underlying lattice and the algebraic structure of the Ising partition
function.

The Kac--Ward representation is closely connected with other exact
formulations of the planar Ising model. Fisher showed that the Ising
partition function can be mapped onto a close-packed dimer problem on
an appropriately decorated planar graph \cite{Fisher-1966}. In this
formulation the partition function is evaluated by Pfaffian methods.
The equivalence between spinor, dimer, Pfaffian, Kac--Ward, and
free-fermion descriptions reflects the underlying fermionic structure
of the two-dimensional Ising model
\cite{SML-1964,Vdovichenko-1964,Popov-book,Polyakov-book,Samuel-1980,Sedrakyan-1984}.

A fermionic field formulation based directly on the local Boltzmann
weights and their representation by graded $R$ operators was developed
in Ref.~\cite{KhS1}. In this approach the partition function of the
2DIM is represented as a Grassmann functional integral with a quadratic
fermionic action. The standard results for the free energy and specific
heat were reproduced within this formulation, and determinant
representations for the spin--spin correlation functions were obtained.
The method also provides a natural extension to the eight-vertex
($XYZ$) model and allows one to formulate the Ising partition function
in the presence of an external magnetic field. This fermionic
$R$-operator formulation is closely related to the approach used in
the present work and provides an important starting point for its
extension to more complicated lattice geometries. In a recent related
study, Khachatryan, Babujian, and Sedrakyan considered a generalized
inhomogeneous kagom\'e Ising model with two alternating couplings along
each lattice direction, mapped its local Boltzmann weights to a
non-symmetric eight-vertex $R$-matrix satisfying the free-fermion
condition, and analyzed the corresponding Yang--Baxter structure and
anisotropic-limit quantum spin chains \cite{Khachatryan-Babujian-Sedrakyan-2026}.
Formulation of action as fermionic field theory for arbitrary network
models with two and multiparticle scattering matrices was formulated in
\cite{KhSSP,KhSSR}.

 Grassmann variables
provide a natural field-theoretical language for this structure:
Gaussian Grassmann integrals transform the combinatorial problem into
the partition function of free lattice fermions.

The fermionic nature of the 2DIM is also manifested directly in its
correlation functions and elementary excitations. Disorder variables
and fermionic operators provide another formulation of this relation
\cite{Kadanoff-Ceva-1971}. In the vicinity of the critical point the
long-wavelength excitation spectrum becomes that of a massless
two-dimensional fermionic field, whereas away from criticality a
nonzero fermionic mass measures the deviation from the critical
surface. This relation between criticality and the appearance of
fermionic zero modes will be particularly useful in the formulation
developed below.

The Kac--Ward construction has also been generalized far beyond the
original square-lattice problem. It can be formulated for arbitrary
finite graphs and for graphs embedded in surfaces of nonzero genus
\cite{Cimasoni-2010}. For critical Ising models on planar graphs,
Kac--Ward matrices exhibit explicit relations with duality and
discrete Laplacians \cite{Cimasoni-2012}. For arbitrary planar
doubly-periodic weighted graphs the critical temperature can be
characterized directly by an algebraic equation in the corresponding
edge weights \cite{Cimasoni-Duminil-2013}. The relations between
Kac--Ward matrices, Grassmann variables, dimers, disorder operators,
and fermionic observables have also been clarified in a general
combinatorial framework \cite{Chelkak-Cimasoni-Kassel-2017}.
These developments demonstrate that the fermionic approach is not
restricted to a particular regular lattice but is naturally adapted
to general planar geometries.

A closely related line of work has emphasized exact critical
temperatures and thermodynamic properties for broad classes of planar
Ising lattices. Codello used the Feynman--Vdovichenko combinatorial
approach to determine the exact Curie temperatures of all Archimedean
lattices and, by duality, of their Laves duals \cite{Codello-2010}.
More recently, Ben Al\`i Zinati and Codello introduced a universal
planar-Ising emulator in which different lattices are represented by
site- or bond-diluted supercells of a fixed host lattice, so that the
underlying Feynman--Vdovichenko/Kac--Ward matrix is kept fixed while
lattice dependence is encoded by binary masks \cite{Codello-2026}.
Pierre, Bernu, and Messio derived general expressions for the free
energy, entropy, and specific heat of planar Ising models and applied
them systematically to Archimedean lattices and their duals
\cite{Pierre-2025}.

Complementary studies have addressed frustration, dynamics, and general
constraints on criticality. Yu systematically investigated the Ising
antiferromagnet on the Archimedean lattices, with particular attention
to geometric frustration, residual entropy, and the character of the
ground states \cite{Yu-2015}; the zero-temperature Glauber dynamics of
ferromagnetic and antiferromagnetic models on Archimedean and
2-uniform lattices was subsequently analyzed in Ref.~\cite{Yu-2017}.
Joseph and Boettcher recently derived exact upper bounds on the critical
temperature of ferromagnetic Ising models on periodic planar
tessellations within the Feynman--Kac--Ward formalism, with the bound
controlled by the maximal coordination number \cite{Joseph-2026-2}.
Related work constructed families of periodic planar lattices with
arbitrarily high critical temperature and studied the corresponding
scaling with the maximal coordination number \cite{Joseph-2026-1}.
These results further illustrate the usefulness of Kac--Ward and
fermionic methods for systematic studies of Ising models on different
planar geometries.

Of particular interest in this respect are the kagom\'e and dice 
lattices. Its
elementary cell contains three sites, and its geometry combines
triangular elementary plaquettes with hexagonal ones. The Ising model
on the kagom\'e lattice was considered already in the early development
of exact lattice statistical mechanics. Sy\^ozi obtained the critical
temperature of the isotropic model \cite{Syozi-1951}, while Kan\^o and
Naya investigated both the ferromagnetic and antiferromagnetic kagom\'e
Ising models \cite{Kano-1953} and demonstrated their solvability.  The spontaneous magnetization was studied
by Naya \cite{Naya-1954}, and various transformations relating kagom\'e,
honeycomb, triangular, and decorated lattices were subsequently
developed \cite{Syozi-1955,Syozi-1960}. The kagom\'e geometry is also
especially important in the antiferromagnetic case, where triangular
plaquettes generate geometric frustration and lead to a macroscopically
degenerate ground-state structure.

More recently, distorted and anisotropic kagom\'e Ising models have
been investigated by exact and numerical methods. Phase diagrams and
thermodynamic properties of distorted kagom\'e lattices were studied
in Ref.~\cite{Li_2010}, while exact free-fermion techniques were also
applied to special kagom\'e models in an external magnetic field
\cite{Lu-Wu-2005}. Questions concerning spontaneous magnetization and
frustration for general signs and magnitudes of the kagom\'e couplings
have continued to attract attention
\cite{Kassan-Ogly-2023,Sedrakyan-2026}. These studies show that the
kagom\'e lattice remains a useful testing ground for exact methods
because it combines a nontrivial unit cell, anisotropy, frustration,
and lattice duality.

The planar dual of the kagom\'e lattice is the dice, or $T_3$, lattice.
Its geometry is qualitatively different from that of the original
kagom\'e lattice: the elementary cell contains sites with coordination
numbers three and six. Ising models and frustrated spin systems on the
dice lattice have also been studied in different contexts
\cite{Valdes-2007}. From the point of view of the present approach,
the kagom\'e--dice pair is particularly interesting because it allows
one to study explicitly how Kramers--Wannier duality is realized at
the level of the fermionic matrices and their spectral determinants.

In the present work we use a single fermionic construction to analyze
the 2DIM on several planar lattices. Grassmann fields are associated
with directed links, and the change of direction of a fermionic
trajectory at each vertex is accompanied by the corresponding
Kac--Ward rotation factor. The resulting action is quadratic, so after
Fourier transformation the partition function is determined by a
finite-dimensional momentum-space matrix. The critical coupling is
then identified with the appearance of a zero mode \cite{Sedrakyan-1984,Cimasoni-2010},
\bea
{\det\cal A}(\bf k=0)=0.
\ena
At the same time, the expansion of ${\det \cal A}(\bf k)$ around zero
momentum gives the long-wavelength fermionic spectrum and allows the
long-wavelength fermionic mass to be determined near criticality. Its vanishing locates the classical thermal transition. Exact lattice decay lengths require the complex zeros of the full spectral polynomial and the appropriate spin insertion.

We first demonstrate the method for the square, honeycomb, and triangular lattices, for which the standard exact critical couplings are reproduced. We then consider the kagom\'e lattice and its dual, the dice lattice, with three independent coupling constants and derive the corresponding anisotropic critical surfaces and spectral equations. In the isotropic limit, the known critical points of the kagom\'e and dice lattices are recovered.

We next analyze the long-wavelength spectra and their vanishing
fermionic masses. For the fully anisotropic kagom\'e model we combine
the fermionic spectral curve with the exact local transformation of the
physical spins. We also determine the magnetizations of both types of
dice sites. Here and below $k_0$ is a Euclidean momentum. A real
transfer-matrix energy requires analytic continuation in a specified
lattice direction; the quadratic momentum expansions are used near
criticality.

In this way, the critical properties of several geometrically different Ising lattices are described within a single fermionic Kac--Ward framework. The results also provide a natural basis for extending the approach to frustrated Ising models on more general lattices. 

\section{Fermionic field representation}

We take spins $s_v=\pm1$ and dimensionless couplings
$J_e=\beta\mathcal J_e>0$, where $\mathcal J_e$ is the exchange energy
and $\beta=(k_{\rm B}T)^{-1}$. There is no external field. Magnetizations
are positive spontaneous values selected by plus boundary conditions
or by taking the thermodynamic limit before $h\to0^+$.
For a finite planar graph with $n_v$ vertices and edge set $E$, the
high-temperature expansion is
\begin{align}
Z&=\sum_{\{s\}}\exp\!\left(\sum_{e=uv}J_e s_us_v\right)
=2^{n_v}\prod_{e\in E}\cosh J_e\,\mathcal P(x),\label{Z}\\
\mathcal P(x)&=\sum_{F\subset E:\,\partial F=\varnothing}
\prod_{e\in F}x_e,\qquad x_e=\tanh J_e.\nonumber
\end{align}
An even subgraph $F$ has even degree at every vertex. In the homogeneous
case its weight is $\lambda^L$, with $\lambda=\tanh J$ and
$L=|F|$; the factor outside the sum is
$2^{n_v}(\cosh J)^{|E|}$.

\begin{figure}[tb]
	\centering
	\includegraphics[width=.7\linewidth]{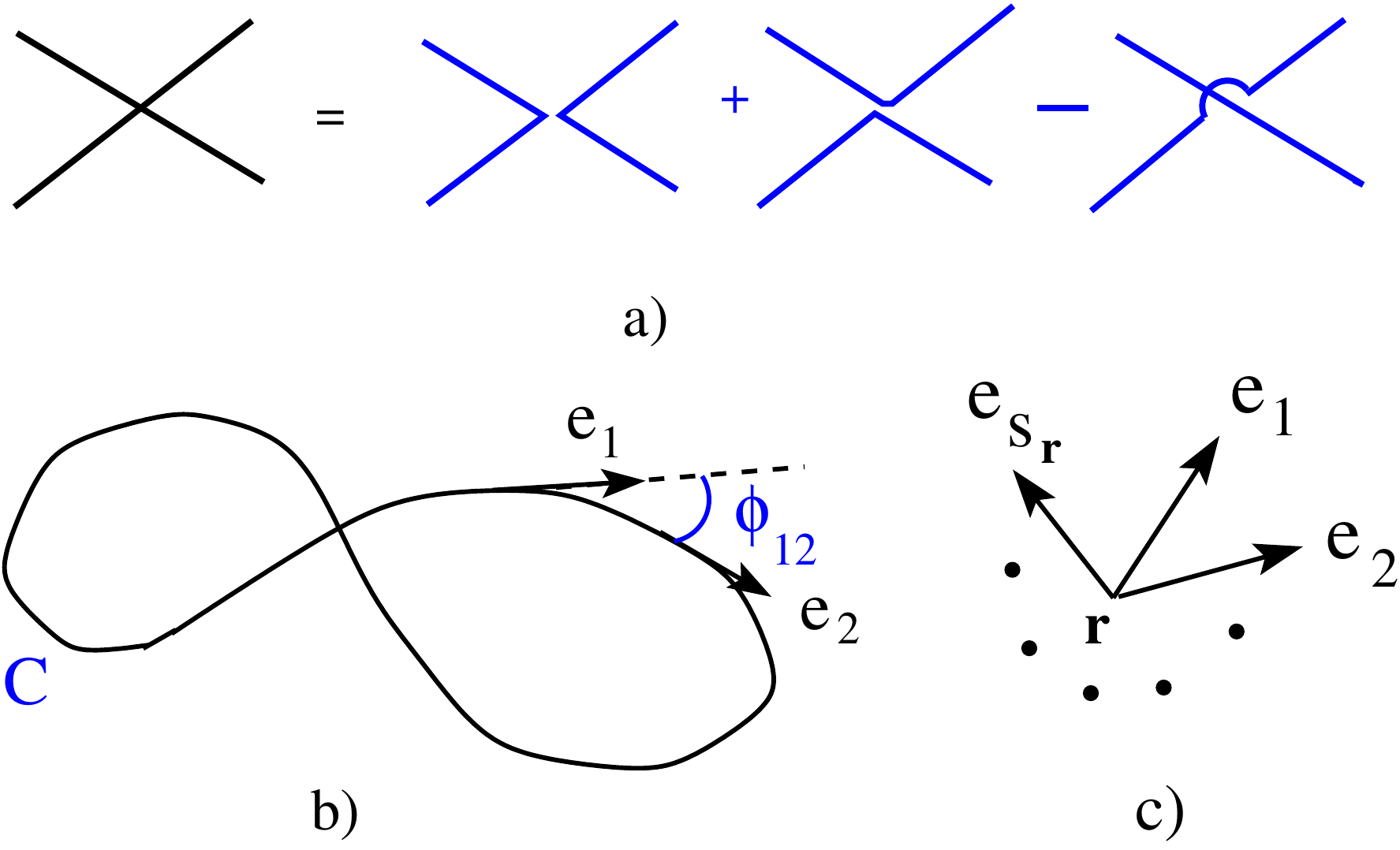}
	\caption{Graphical representation of the Kac--Ward construction. (a) Resolution of a self-intersection into the possible nonintersecting trajectories; the relative signs ensure cancellation of the overcounted configurations. (b) A closed trajectory $C$ with two successive tangent vectors $\vec e_1$ and $\vec e_2$ forming the oriented turning angle $\phi_{12}$, which determines the Kac--Ward phase factor. (c) Directed bonds $\vec e_1,\ldots,\vec e_{s_{\vec r}}$ attached to a lattice site $\vec r$ of coordination number $s_{\vec r}$ and the corresponding fermionic degrees of freedom.}
	\label{self}
\end{figure}

If these loops could be interpreted as particle trajectories, each parameterized by a single coordinate, then, following Feynman's path-integral formulation, $-\log(\tanh J)$ is the Euclidean action per unit length in the convention $e^{-S_E}$ of a bosonic particle. This interpretation, however, is obstructed by the presence of self-intersections. At a self-intersection, a loop can be decomposed into trajectories in several inequivalent ways (see Fig.~\ref{self}a)), so the graphical expansion cannot be identified with a sum over uniquely defined particle worldlines.

The Kac--Ward transition matrix $T$ acts on directed edges. It
allows successive edges only when they join without an immediate
reversal and assigns the phase $e^{i\phi/2}$ for the oriented turning
angle $\phi$. The precise closed-trajectory expansion is
\begin{equation}
\log\mathcal P(x)=\frac12\log\det(I-T)
=-\frac12\sum_{n\ge1}\frac{\operatorname{Tr}T^n}{n}.
\label{Z2}
\end{equation}
The series first holds at sufficiently small couplings; the determinant
identity extends as a polynomial identity. Its factors $1/(2n)$ count
orientations and starting points. Geometrically, the turning phases
cancel the extra resolutions of self-intersections. For a generic
closed planar trajectory with $n$ transverse self-intersections,
the sign is $(-1)^n$.
{The Kac--Ward (KW) factor can be represented as an integral along the trajectory}
\bea
\label{KW-1}
(-1)^n=-e^{\frac{i}{2}\int d\phi}=-e^{\frac{i}{2}\int_0^T \frac{d\phi}{dt} dt}
\ena
and it is connected with the homotopy group $\pi_1(SO(3))$ \cite{Sedrakyan-1984}.
This expression can be written in terms of tangent vectors to the curve $C$.
{Namely, let us define \cite{Sedrakyan-1984}}
\bea
\label{Lambda}
\Lambda_{e_1}^{e_2}=\frac{1+\hat{e_1}\hat{e_2}}{\sqrt{2(1+{\vec e_1}{\vec, e_2})}} = e^{\frac{i}{2} \sigma^z \phi_{12}}
\ena
where $\hat{e}={\vec e}{\vec \sigma},$  ${\vec \sigma}$ are Pauli matrices and
$\phi_{12}$ is an angle between vectors $\vec e_1$ and $\vec e_2$ (see Fig.~\ref{self}b). Then, it is straightforward to see that
\bea\label{KW-2}
\hspace{-0.5cm}(-1)^n=-\frac{1}{2} Tr\Big[\prod_{i=1}^L \Lambda_{e_i}^{e_{i+1}}\Big]=-e^{\frac{i}{2} \sum_{i=1}^L \phi_{i,i+1}}=e^{i \pi n} 
\ena 
{Figures~2(a)--5(a) illustrate how this phase factor is generated on the regular, honeycomb, triangular, and kagom\'e lattices, respectively. A natural question is whether the Kac--Ward factor can be interpreted as the action of a bosonic or fermionic field. Such a representation transforms the combinatorial loop expansion of the Ising model into a genuine path-integral formulation. This problem was solved by Vdovichenko \cite{Vdovichenko-1964} (see also \cite{Popov-book,Polyakov-book,Sedrakyan-1984}), who demonstrated that the two-dimensional Ising model is equivalent to a theory of free lattice fermions. His construction showed that the KW phase factor naturally arises from the propagation of fermionic degrees of freedom.}

{The presence of fermions in the 2DIM is evident from the sign factor, which ensures the cancellation of self-intersecting trajectories as a consequence of the Pauli principle.}

To construct the action on lattices with different coordination numbers,
label a directed edge by $e$ and its reversal by $\bar e$. The transition
matrix is
\begin{equation}
T_{ee'}=\begin{cases}
x_e e^{i\phi(e,e')/2},&\mathrm{head}(e)=\mathrm{tail}(e'),\ e'\ne\bar e,\\
0,&\text{otherwise},
\end{cases}
\end{equation}
where $x_e=\tanh J_e$ and $-\pi<\phi<\pi$. In one eigen-sector of
$\Lambda$, the action is
\begin{equation}
S(\psi)=\sum_{e,e'}\bar\psi_e
\bigl(T_{ee'}-\delta_{ee'}\bigr)\psi_{e'}.
\label{S1}
\end{equation}
Each directed state has one local diagonal term. With the Grassmann
measure fixed by the determinant convention, the planar identity is
\begin{equation}
\mathcal P(x)^2=\det(I-T)
=\int\mathcal D\bar\psi\mathcal D\psi\,e^{S(\psi)}.
\label{Z3}
\end{equation}
Equation~(\ref{Z}) restores the physical partition function. On a finite
torus the physical partition function combines four fermionic boundary
sectors \cite{Cimasoni-2010,Chelkak-Cimasoni-Kassel-2017}. We define spin
correlations first on finite planar domains, then take their
thermodynamic limit at fixed separation.

The normalized spin correlation and spontaneous magnetization are
\begin{align}
\langle s_as_b\rangle
&=\frac{1}{Z}\sum_{\{s\}}s_as_b
e^{\sum_{e=uv}J_e s_us_v},\nonumber\\
M_a^2&=\lim_{|a-b|\to\infty}\langle s_as_b\rangle,
\label{M1}
\end{align}
with the same sublattice at both ends of the last limit. For the action
$S=\bar\psi\mathcal A\psi$, we use
\begin{equation}
G_{\mu\nu}(R)=\langle\bar\psi_{\nu,0}\psi_{\mu,R}\rangle,
\qquad G(\mathbf k)=\mathcal A^{-1}(\mathbf k).
\label{G}
\end{equation}
For a unit cell containing several inequivalent sites, the fermionic
indices are composite indices
\[
\mu=(i,\alpha),\qquad \nu=(j,\beta).
\]
where $i,j$ label the sites inside the unit cell and $\alpha,\beta$
label the directed-link fermionic components attached to those sites.
Thus $G_{(i,\alpha)(j,\beta)}(R)$ is the propagator from directed
component $(j,\beta)$ in the reference cell to component $(i,\alpha)$
in the cell displaced by $R$. For a one-site unit cell the site index is
trivial and only the directed-link index remains.
All Fourier integrals use the unit-cell measure
$d\mu(\mathbf k)=A_{\rm cell}d^2k/(2\pi)^2$, so that
$\int_{\rm BZ}d\mu=1$ in Cartesian reciprocal coordinates.

The spin insertion requires a path. For a simple path $\gamma$ of $m$
edges joining $a$ to $b$, let $P_\gamma$ select both directed states of
every path edge in the canonical order specified in
Appendix~\ref{app:path-pfaffian}. Put
\begin{equation}
J_2=\begin{pmatrix}0&1\\-1&0\end{pmatrix},\qquad
D_\gamma=\bigoplus_{e\in\gamma}\sqrt{x_e^{-1}-x_e}\,I_2.
\end{equation}
The explicit $2m\times2m$ correlation matrix is
\begin{align}
\Gamma_\gamma
&=\bigoplus_{e\in\gamma}x_e^{-1}J_2
+D_\gamma P_\gamma^T\widehat K^{-1}P_\gamma D_\gamma,
\label{path-Pf-main}\\
\langle s_as_b\rangle&=\operatorname{Pf}\Gamma_\gamma.
\label{spin-path-pfaffian}
\end{align}
The matrix $\Gamma_\gamma$ is the $2m\times2m$ real antisymmetric
correlation matrix associated with the spin insertion along the path
$\gamma$. Its first term contains the local contact blocks on the path,
whereas the second term is the fermionic propagator projected onto the
$2m$ directed states selected by $P_\gamma$. More explicitly,
Eq.~(\ref{eq:green-to-real-skew}) gives
\[
\widehat K^{-1}
=i\mathsf U^\dagger\mathsf W^{-1}
\mathcal A^{-1}\mathsf W\mathsf J\mathsf U,
\]
and therefore
\bea
\Gamma_\gamma
&=&Q_\gamma
+iD_\gamma P_\gamma^T\mathsf U^\dagger\mathsf W^{-1}
\mathcal A^{-1}\mathsf W\mathsf J\mathsf U
P_\gamma D_\gamma,\nn\\
Q_\gamma &\equiv&\bigoplus_{e\in\gamma}x_e^{-1}J_2.
\ena
Since $\mathcal A^{-1}(\mathbf k)=G(\mathbf k)$, every nonlocal entry of
$\Gamma_\gamma$ is obtained from the real-space Green functions
$G_{\mu\nu}(R)$ of Eq.~(\ref{G}), with the appropriate unit-cell
separation between the directed states selected by $P_\gamma$.
Thus $\Gamma_\gamma$ is constructed from the fermionic Green function,
but it also contains the contact blocks and path weights required for
the physical spin correlator.
Here $\widehat K$ is the real antisymmetric Kac--Ward matrix in
Eq.~(\ref{eq:kw-real-skew}); Eq.~(\ref{eq:green-to-real-skew}) expresses
its inverse through $G$. Appendix~\ref{app:path-pfaffian} proves
Eq.~(\ref{spin-path-pfaffian}), including its sign, contact blocks, edge
weights, and path independence. It also performs the exact
four-index endpoint Schur reduction and keeps its interior Pfaffian.
The subsequent local Gaussian calculation in
Appendix~\ref{app:local-schur} determines the physical spin factors.
With the composite-index notation defined above, the physical spin correlator is
the full path Pfaffian $\operatorname{Pf}\Gamma_R$, not a single
$G_{\mu\nu}$ or a fixed finite sum of Green functions. Because the
Pfaffian dimension grows with $R$, its ordered-phase limit may remain
nonzero although each fermionic Green function is short ranged.

We will use now the action (\ref{S1}) for various lattices to calculate
analytically precisely the value of critical coupling of the 2DIM on that lattice.

\section{Critical coupling of 2DIM on regular lattice}

\begin{figure}[tb]
	\centering
	\includegraphics[width=.6\linewidth]{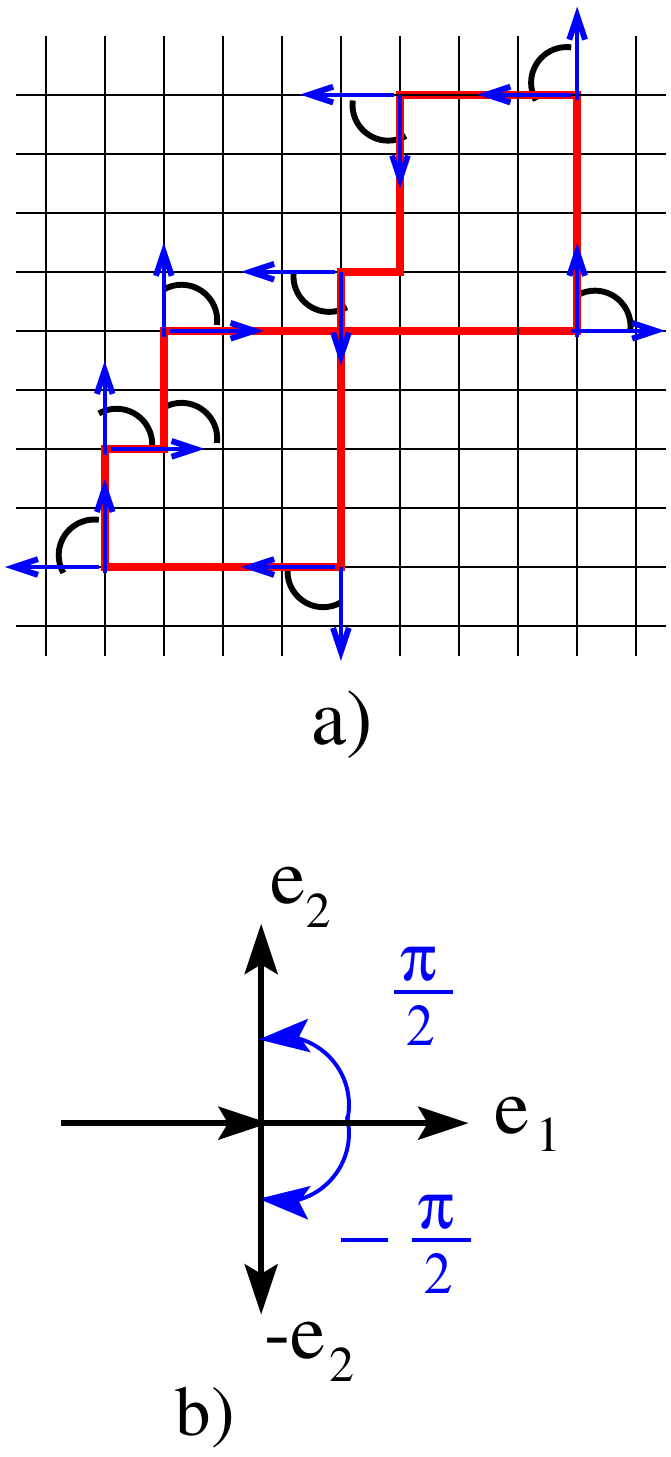}
\caption{(a) Formation of the Kac--Ward (KW) phase factor on the regular
	(quadrangular) lattice. (b) Elementary unit vectors and the corresponding
	turning angles between the directed lattice bonds.}
	\label{regular}
\end{figure}

{The unit cell of the regular lattice contains one site with coordination number four.} We attach to each site ${\vec r}$ four fermions $\psi^{\pm e_\alpha}_{\vec r}, \alpha =1,2$ associated with vectors $\pm e_1, \pm e_2$, see Fig.\ref{regular}b. 

According to the expression (\ref{Lambda}) here $\Lambda_{e_1}^{e_2}=e^{i\pi/4 }$,
$\Lambda_{e_1}^{e_1}=1$ and $\Lambda_{e_1}^{-e_2}=e^{-i\pi/4 }$. Since regular lattice is translational invariant we can make Fourier transform and pass to momentum space. By introducing four component vector 
$\Psi_{\bf k}=(\psi^{e_1}_{\bf k},\psi^{-e_1}_{\bf k},\psi^{e_2}_{\bf k},\psi^{-e_2}_{\bf k})$ with
\bea
\label{Fourier}
\psi_{\vec r+\vec e_i}^{e_j}
=
\frac{1}{\sqrt{N}}
\sum_{\vec k}
e^{i\vec k\cdot\vec r}
e^{i\vec k\cdot\vec e_i}
\psi_{\vec k}^{e_j}
\ena
we can write an
action (\ref{S1}) in a form
\bea
\label{S2}
	S(\Psi)=
\sum_{\vec k}\Psi_{\bf k}^\dagger {\cal A}(\vec k) \Psi_{\bf k},
\ena
%
where the matrix ${\cal A}(\vec k)$, following directly from
Eq.~(\ref{S1}), is
%
\bea
{\cal A}(\vec k)= \hspace{7.5cm} \\
\left(
\begin{array}{cccc}
	-1+\lambda e^{ik_1}
	&
	0
	&
	\lambda e^{ik_1}\Lambda_{e_1}^{e_2}
	&
	\lambda e^{ik_1}\Lambda_{e_1}^{-e_2}
	\\[2mm]
	0
	&
	-1+\lambda e^{-ik_1}
	&
	\lambda e^{-ik_1}\Lambda_{-e_1}^{e_2}
	&
	\lambda e^{-ik_1}\Lambda_{-e_1}^{-e_2}
	\\[2mm]
	\lambda e^{ik_2}\Lambda_{e_2}^{e_1}
	&
	\lambda e^{ik_2}\Lambda_{e_2}^{-e_1}
	&
	-1+\lambda e^{ik_2}
	&
	0
	\\[2mm]
	\lambda e^{-ik_2}\Lambda_{-e_2}^{e_1}
	&
	\lambda e^{-ik_2}\Lambda_{-e_2}^{-e_1}
	&
	0
	&
	-1+\lambda e^{-ik_2}
\end{array}
\right),\nn
\label{A-matrix-general}
\ena
where  the conditions 
$ \Lambda_{e_1}^{e_1}=\Lambda_{-e_1}^{-e_1}=\Lambda_{e_2}^{e_2}=
\Lambda_{-e_2}^{-e_2}=1 $ were used,
since propagation along the same direction does not involve a
rotation of the tangent vector.

The vanishing matrix elements correspond to immediate
backtracking processes,$
e_1\rightarrow -e_1,\;
-e_1\rightarrow e_1,\;
e_2\rightarrow -e_2,\;
-e_2\rightarrow e_2,
$ which are excluded in the Kac--Ward representation.

The matrix ${\cal A}(k)$ defines the dynamics of Ising model. Its determinant 
\begin{align}
\det\mathcal A(k)&=(1+\lambda^2)^2\nonumber\\
&\quad-2\lambda(1-\lambda^2)(\cos k_0+\cos k).
\label{det}
\end{align}  
where $k_0=k_1$ and $k=k_2$ are Euclidean momentum coordinates. The condition $\det A(\vec k)=0$ defines the spectral curve. A direct calculation gives
\begin{align}
\cos k_0+\cos k
&=\frac{(1+\lambda^2)^2}{2\lambda(1-\lambda^2)}\nonumber\\
&=\sinh2J+\frac{1}{\sinh2J}.
\label{zeroA}
\end{align}
where we took into account that $\lambda=\tanh J$. 
For $J>0$ the right-hand side is not smaller than $2$,
while $\cos k_0+\cos k\leq 2$. {Expanding the cosine terms around $k_0=k=0$, one obtains} 
\bea
\label{spec-1}
k_0^2+k^2=4-2\sinh 2J-2\frac{1}{\sinh 2J}=-m^2
\ena
Zero mode
appears at $k_0=k=0,\qquad \sinh 2J_c=1,$
which yields
\bea
J_c=\frac{1}{2}\ln(1+\sqrt{2}),
\quad
\lambda_c=\sqrt{2}-1=\frac{1}{1+\sqrt{2}}.
\ena



  \begin{figure}[tb]
 	\centering
 	\includegraphics[width=.6\linewidth]{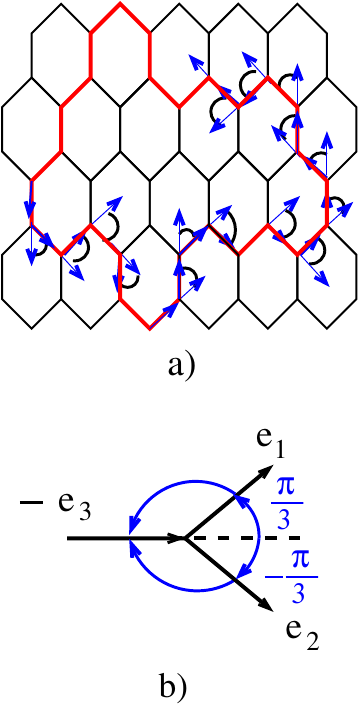}
 	\caption{Fermionic trajectories and Kac--Ward rotations on the honeycomb lattice. (a) An example of an oriented fermionic trajectory on the honeycomb lattice. (b) Elementary rotations of the fermionic tangent vector at a vertex of coordination three. The incoming direction $-\vec e_3$ turns into either outgoing direction $\vec e_1$ or $\vec e_2$, giving angles $\pm\pi/3$; the outgoing bond vectors satisfy $\vec e_1+\vec e_2+\vec e_3=0$.}
 	\label{honeycomb}
 \end{figure}


 \section{Critical coupling of the 2DIM on the honeycomb lattice}
 
 We apply the fermionic representation (\ref{S1}) to the
 two-dimensional Ising model on the honeycomb lattice. Since every
 site of the honeycomb lattice has coordination number three, three
 fermionic variables associated with the three directed links are
 attached to each lattice site.
 
According to Fig.~3b, each lattice site has coordination number
three. We denote the three directed vectors at one sublattice by
$ e_1,\; e_2,\; e_3,$
as shown in Fig.~3b. With this convention the three vectors satisfy
\bea
\vec{e}_1+\vec{e}_2+\vec{e}_3=0.
\label{honey-vector-relation}
\ena
The opposite sublattice is characterized by the reversed vectors
$-\vec{e}_1,\; -\vec{e}_2,\; -\vec{e}_3.$
We therefore introduce the three-component fermionic fields
\bea
\Psi_A(\vec k)
&=&
\left(
\psi_{\vec k}^{e_1},
\psi_{\vec k}^{e_2},
\psi_{\vec k}^{e_3}
\right)^T, \nn\\
\Psi_B(\vec k)
&=&
\left(
\psi_{\vec k}^{-e_1},
\psi_{\vec k}^{-e_2},
\psi_{\vec k}^{-e_3}
\right)^T.
\label{honey-fields}
\ena
Starting from the fermionic action (\ref{S1})
and performing the Fourier transformation
\bea
\psi_{\vec r}^{\,e_i}
=
\frac{1}{\sqrt N}
\sum_{\vec k}
e^{i\vec k\cdot\vec r}
\psi_{\vec k}^{\,e_i},
\ena
the action can be represented in the form
\bea
S=
\sum_{\vec k}
\left(
\bar\Psi_A,\bar\Psi_B
\right)
{\cal A}(\vec k)
\left(
\begin{array}{c}
	\Psi_A\\
	\Psi_B
\end{array}
\right),
\ena
where
\bea
{\cal A}(\vec k)
=
\left(
\begin{array}{cc}
	-I_3 & \lambda C(\vec k)\\
	\lambda D(\vec k)&-I_3
\end{array}
\right).
\label{honey-A6}
\ena
The matrix describing propagation from the $A$ to the $B$
sublattice is
\bea
C(\vec k)=
\left(
\begin{array}{ccc}
	0
	&
	e^{i\vec k\cdot \vec{e}_1}\Lambda_{e_1}^{-e_2}
	&
	e^{i\vec k\cdot \vec{e}_1}\Lambda_{e_1}^{-e_3}
	\\[2mm]
	e^{i\vec k\cdot \vec{e}_2}\Lambda_{e_2}^{-e_1}
	&
	0
	&
	e^{i\vec k\cdot \vec{e}_2}\Lambda_{e_2}^{-e_3}
	\\[2mm]
	e^{i\vec k\cdot \vec{e}_3}\Lambda_{e_3}^{-e_1}
	&
	e^{i\vec k\cdot \vec{e}_3}\Lambda_{e_3}^{-e_2}
	&
	0
\end{array}
\right),\qquad
\label{honey-C}
\ena
while propagation in the opposite direction is described by
\bea
D(\vec k)=
\left(
\begin{array}{ccc}
	0
	&
	e^{-i\vec k\cdot \vec{e}_1}\Lambda_{-e_1}^{e_2}
	&
	e^{-i\vec k\cdot \vec{e}_1}\Lambda_{-e_1}^{e_3}
	\\[2mm]
	e^{-i\vec k\cdot \vec{e}_2}\Lambda_{-e_2}^{e_1}
	&
	0
	&
	e^{-i\vec k\cdot \vec{e}_2}\Lambda_{-e_2}^{e_3}
	\\[2mm]
	e^{-i\vec k\cdot \vec{e}_3}\Lambda_{-e_3}^{e_1}
	&
	e^{-i\vec k\cdot \vec{e}_3}\Lambda_{-e_3}^{e_2}
	&
	0
\end{array}
\right).\qquad
\label{honey-D}
\ena
The vanishing diagonal elements correspond to immediate
backtracking, which is excluded in the Kac--Ward construction.

Since the honeycomb lattice is bipartite, one of the two
sublattices can be eliminated. The resulting effective
three-component action is
\bea
S_{\rm eff}
=-\sum_{\vec k}
\bar\Psi_A(\vec k)A(\vec k)\Psi_A(\vec k),
\ena
with
\bea
	A(\vec k)=I_3-\lambda^2 C(\vec k)D(\vec k).
\label{honey-A3}
\ena
The critical point corresponds to the appearance of a zero mode
of this matrix and is therefore determined by
\bea
\det A(\vec k)=0.
\label{honey-critical-condition}
\ena
In calculating the determinant we use the relations
\bea
\Lambda_{e_i}^{e_j}\Lambda_{e_j}^{e_i}=1.
\label{honey-Lambda-rel}
\ena
The remaining products of the $\Lambda$ matrices correspond to
closed fermionic trajectories and produce the Kac--Ward factor
associated with the complete rotation of the tangent vector.
After using these relations, the determinant reduces to
\bea
&&\det A(\vec k)
=
1+3\lambda^4
-2\lambda^2(1-\lambda^2)
\left(
\cos[\vec k\cdot(\vec{e}_1-\vec{e}_2)]
\right.
\nonumber\\
&&\left.
+\cos[\vec k\cdot(\vec{e}_1-\vec{e}_3)]
+\cos[\vec k\cdot(\vec{e}_2-\vec{e}_3)]
\right).
\label{honey-det-vector}
\ena
Thus, in the present convention, the determinant has a completely
symmetric form with respect to the three vectors
$e_1,e_2,e_3$.

Introducing
$k_1=\vec k\cdot \vec{e}_1,\;
k_2=\vec k\cdot \vec{e}_2,\;
k_3=\vec k\cdot \vec{e}_3,$
the geometrical relation (\ref{honey-vector-relation}) implies
\bea
k_1+k_2+k_3=0.
\label{k-relation}
\ena
The determinant can therefore be written in the particularly
symmetric form
\bea
&&\det A(\vec k)
	=
	1+3\lambda^4
	-2\lambda^2(1-\lambda^2)\left(
	\cos[k_1-k_2]\right. \nn\\
&&\left.	+\cos[k_1-k_3]
	+\cos[k_2-k_3]
	\right).
\label{honey-det-k123}
\ena
Only two of the three momentum variables are independent. Defining  $k_0=(k_1-k_2)/\sqrt{3}$  as
energy in Euclidean space, $k=k_1+k_2$ as momentum and $k_3=-k_1-k_2=-k$
%
%
one can alternatively write
\bea
&&\det A(\vec k)
=
1+3\lambda^4
-2\lambda^2(1-\lambda^2)
\Bigg(
\cos[\sqrt{3} k_0]\nn\\
&&+\cos\left[\frac{3 k+\sqrt{3} k_0}{2}\right]
+\cos\left[\frac{3 k-\sqrt{3} k_0}{2}\right]
\Bigg).
\label{honey-det-k12}
\ena

For the ferromagnetic model the zero of the determinant first
appears at zero complex energy and momentum $k_0=k=0 $.

At this point one can see, that
\bea
	\det A(0)=\left(1-3\lambda^2\right)^2
\label{honey-det-zero}
\ena
and the critical value of $\lambda$ is consequently determined as
\bea
	\lambda_c=\tanh[J_c]=\frac{1}{\sqrt3}, \quad
\label{honey-lambda-critical}
	J_c=\frac{1}{2}\ln(2+\sqrt3).
\ena
Thus, the fermionic representation based on the symmetric set of
three vectors $e_1+e_2+e_3=0$
reproduces the exact critical coupling of the two-dimensional
Ising model on the honeycomb lattice.

Finally, the spectral equation of 2DIM on honeycomb lattice at low energies and momentum follows from (\ref{honey-det-k12}) and reads
\bea
\label{spec-4}
k_0^2+k^2 +m^2=0,\; m^2=\frac{2(1-3\lambda^2)^2}{9 \lambda^2(1-\lambda^2)}
\ena
 
 \section{Critical coupling of 2DIM on the triangular lattice} 

\begin{figure}[tb]
	\centering
	\includegraphics[width=0.6\linewidth]{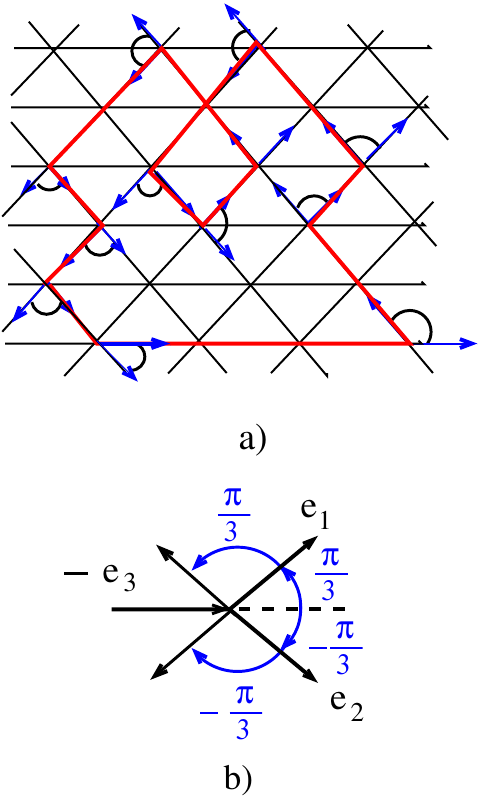}
	\caption{Fermionic trajectories and Kac--Ward rotations on the triangular lattice. (a) An example of an oriented fermionic trajectory on the triangular lattice. (b) Elementary rotations of the fermionic tangent vector at a lattice site. The three elementary directions $\vec e_1$, $\vec e_2$, and $\vec e_3$ are chosen according to the symmetric convention $\vec e_1+\vec e_2+\vec e_3=0$, and the corresponding turning angles are multiples of $\pi/3$.}
	\label{triangular}
\end{figure}

We consider the isotropic Ising model on the triangular lattice
shown in Fig.~\ref{triangular}.  It is convenient to choose the
three elementary vectors in the symmetric convention (see Fig.3b)
$
\vec{e}_1+\vec{e}_2+\vec{e}_3=0.
$
At each lattice site there are six directed bonds,
$ e_1,\quad e_2,\quad e_3,\quad
-e_1,\quad -e_2,\quad -e_3 $.
Introducing
\bea
k_i=\vec k\cdot e_i ,
\ena
the geometrical relation between the vectors implies
$
k_1+k_2+k_3=0,
\qquad
k_3=-k_1-k_2 .
$

We start from the fermionic action (\ref{S1}).
By making Fourier transformation (\ref{Fourier})
and introducing  six-component field
\bea
\label{psi-6}
\Psi_{\vec k}
=
\left(
\psi_{\vec k}^{e_1},
\psi_{\vec k}^{e_2},
\psi_{\vec k}^{e_3},
\psi_{\vec k}^{-e_1},
\psi_{\vec k}^{-e_2},
\psi_{\vec k}^{-e_3}
\right)^T 
\ena
{we obtain the action in the form} 
\bea
S=
\sum_{\vec k}
\bar\Psi_{\vec k}\,
{\cal A}(\vec k)\,
\Psi_{\vec k}.
\ena
In the basis chosen above the momentum-space fermionic operator is

\begin{widetext}
	\bea
\resizebox{0.9\textwidth}{!}{$
	{\cal A}(\vec k)
	=
	\left(
	\begin{array}{cccccc}
		-1+\lambda e^{ik_1}
		&
		\lambda e^{ik_1}\Lambda_{e_1}^{e_2}
		&
		\lambda e^{ik_1}\Lambda_{e_1}^{e_3}
		&
		0
		&
		\lambda e^{ik_1}\Lambda_{e_1}^{-e_2}
		&
		\lambda e^{ik_1}\Lambda_{e_1}^{-e_3}
		\\[1mm]
		\lambda e^{ik_2}\Lambda_{e_2}^{e_1}
		&
		-1+\lambda e^{ik_2}
		&
		\lambda e^{ik_2}\Lambda_{e_2}^{e_3}
		&
		\lambda e^{ik_2}\Lambda_{e_2}^{-e_1}
		&
		0
		&
		\lambda e^{ik_2}\Lambda_{e_2}^{-e_3}
		\\[1mm]
		\lambda e^{ik_3}\Lambda_{e_3}^{e_1}
		&
		\lambda e^{ik_3}\Lambda_{e_3}^{e_2}
		&
		-1+\lambda e^{ik_3}
		&
		\lambda e^{ik_3}\Lambda_{e_3}^{-e_1}
		&
		\lambda e^{ik_3}\Lambda_{e_3}^{-e_2}
		&
		0
		\\[1mm]
		0
		&
		\lambda e^{-ik_1}\Lambda_{-e_1}^{e_2}
		&
		\lambda e^{-ik_1}\Lambda_{-e_1}^{e_3}
		&
		-1+\lambda e^{-ik_1}
		&
		\lambda e^{-ik_1}\Lambda_{-e_1}^{-e_2}
		&
		\lambda e^{-ik_1}\Lambda_{-e_1}^{-e_3}
		\\[1mm]
		\lambda e^{-ik_2}\Lambda_{-e_2}^{e_1}
		&
		0
		&
		\lambda e^{-ik_2}\Lambda_{-e_2}^{e_3}
		&
		\lambda e^{-ik_2}\Lambda_{-e_2}^{-e_1}
		&
		-1+\lambda e^{-ik_2}
		&
		\lambda e^{-ik_2}\Lambda_{-e_2}^{-e_3}
		\\[1mm]
		\lambda e^{-ik_3}\Lambda_{-e_3}^{e_1}
		&
		\lambda e^{-ik_3}\Lambda_{-e_3}^{e_2}
		&
		0
		&
		\lambda e^{-ik_3}\Lambda_{-e_3}^{-e_1}
		&
		\lambda e^{-ik_3}\Lambda_{-e_3}^{-e_2}
		&
		-1+\lambda e^{-ik_3}
	\end{array}
	\right).
	\label{triangular-A} $}
	\ena
\end{widetext}
The zero matrix elements correspond to immediate reversal of the
direction of propagation. Such backtracking processes are excluded
from the Kac--Ward trajectories.

In the present convention the three positive vectors
$e_1,e_2,e_3$ are separated pairwise by an angle $2\pi/3$.
The non-backtracking rotations are therefore
$0,\pm\pi/3,\pm2\pi/3$.  In a fixed eigen-sector of $\sigma_z$
the Kac--Ward factor can be written as
\bea
\Lambda_{e_i}^{e_j}
=
\exp\left(\frac{i}{2}\phi_{ij}\right),
\ena
where $\phi_{ij}$ is the oriented angle between the initial and
final directions.

Because of the relation $k_1+k_2+k_3=0$, the momentum dependence
of the determinant is conveniently expressed through the symmetric
combination
\bea
C(\vec k)
=
\cos k_1+\cos k_2+\cos k_3.
\label{triangular-C}
\ena
Equivalently,
\bea
C(\vec k)
=
\cos k_1+\cos k_2+\cos(k_1+k_2),
\ena
since $k_3=-k_1-k_2$ and the cosine is an even function.

Direct evaluation of the determinant of
Eq.~(\ref{triangular-A}) gives
\bea
	\det{\cal A}(\vec k)
	=
	(1+\lambda^2)^3+8\lambda^3
	-
	2\lambda(1-\lambda^2)^2 C(\vec k).\quad
\label{triangular-det}
\ena
The same result can be written explicitly as a polynomial in
$\lambda$,
\bea
\det{\cal A}(\vec k)
&=&
1-2C(\vec k)\lambda+3\lambda^2
+4(C(\vec k)+2)\lambda^3
\nonumber\\
&+&3\lambda^4-2C(\vec k)\lambda^5+\lambda^6,
\label{triangular-det-expanded}
\ena
The zero-mode condition is
\bea
\det{\cal A}(\vec k)=0,
\ena
which yields
\bea
	\cos k_1+\cos k_2+\cos k_3
	=
	\frac{
		(1+\lambda^2)^3+8\lambda^3
	}{
		2\lambda(1-\lambda^2)^2
	}.
\label{triangular-zero}
\ena
Using $k_3=-k_1-k_2$, this can alternatively be written as
\bea
\cos[k_1]+\cos[k_2]+\cos[k_1+k_2]
=
\frac{
	(1+\lambda^2)^3+8\lambda^3}{2\lambda(1-\lambda^2)^2}.\quad
\ena
Defining Euclidean momentum coordinates $k_0=(k_1-k_2)/\sqrt{3}$ and $k=k_1+k_2$, as for the honeycomb lattice, gives
 following spectral equation
\bea
\label{spec-3}
2 \cos\Big[\frac{k}{2}\Big] \cos\Big[\frac{\sqrt{3} k0}{2}\Big]+\cos[k]
=
\frac{
	(1+\lambda^2)^3+8\lambda^3}{2\lambda(1-\lambda^2)^2}.\qquad
\ena

In order to determine the critical coupling from the closing of the
fermionic gap we notice
\bea
C(\vec k)
=
\cos k_1+\cos k_2+\cos k_3
\leq3.
\ena
The maximal value is attained at the center of the Brillouin zone,
\bea
k_1=k_2=k_3=0.
\ena
In the ferromagnetic region $0<\lambda<1$, the coefficient of
$C(\vec k)$ in Eq.~(\ref{triangular-det}) is negative. Therefore
the determinant reaches zero first at $\vec k=0$.

At zero momentum one obtains
\bea
\label{det-4}
	\det{\cal A}(0)
	=
	(1+\lambda)^2
	(\lambda^2-4\lambda+1)^2 .
\ena
The factor $(1+\lambda)$ does not vanish in the ferromagnetic
region. Hence the critical point is determined by
\bea
\lambda_c^2-4\lambda_c+1=0.
\ena
The two formal solutions are
\bea
\lambda_c=2\pm\sqrt3.
\ena
Since $\lambda=\tanh J$ must satisfy $0<\lambda<1$, the physical
root is
\bea
	\lambda_c=2-\sqrt3
	=
	\frac{1}{2+\sqrt3},\nn\\
%
%
J_c
=
\operatorname{arctanh}(2-\sqrt3)
=
	\frac14\ln3. 
\label{triangular-Jc}
\ena
The critical condition can also be expressed in the standard form.
Using
\bea
\sinh 2J
=
\frac{2\tanh J}{1-\tanh^2J}
=
\frac{2\lambda}{1-\lambda^2},
\ena
we finally obtain
\bea
	\sinh(2J_c)=\frac{1}{\sqrt3}.
\label{triangular-critical}
\ena
The spectral equation of 2DIM on triangular lattice at low energies and momentum follows from (\ref{spec-3}) and reads
\bea
\label{spec-41}
k_0^2+ k^2 + m(\lambda)^2=0,\;\; m(\lambda)^2=\frac{2(\lambda^2-4 \lambda+1)^2}{3 \lambda(1-\lambda)^2} \qquad
\ena

\section{Critical Surface and Spontaneous Magnetization of the Fully Anisotropic Ising Model on the kagom\'e Lattice}

We consider now the fermionic representation of the Ising model
on the kagom\'e lattice. The elementary unit cell contains three
inequivalent sites, which we denote by $i=1,2,3$ (see Fig.5 c)). Each lattice
site has coordination number four and, correspondingly, four
fermionic variables are associated with the four directed links
attached to it.
As in previous lattices and following Fig.5 b), we choose $\vec{e}_1+\vec{e}_2+\vec{e}_3=0.$ for symmetry reasons.  
The four directed states attached to the three sites of the unit
cell can then be represented as
\bea
1:\; \{\pm e_2,\pm e_3\},\;
2:\; \{\pm e_3,\pm e_1\},\;
3:\; \{\pm e_1,\pm e_2\}.\;
\ena
The fermionic hopping part of the action follows directly from
Eq.~(\ref{S1}) written on kagom\'e lattice.

For hopping weights $\lambda_i=\tanh J_i$ along the three
directions, the action is Eq.~(\ref{S1}) with the edge-dependent
transition matrix $T_K$:
\begin{equation}
S(\psi)=\sum_{e,e'}\bar\psi_e
\bigl[(T_K)_{ee'}-\delta_{ee'}\bigr]\psi_{e'}.
\label{S11}
\end{equation}
Thus the local term is counted once for every directed state.
 
\begin{figure}[tb]
	\centering
	\includegraphics[width=0.9\linewidth]{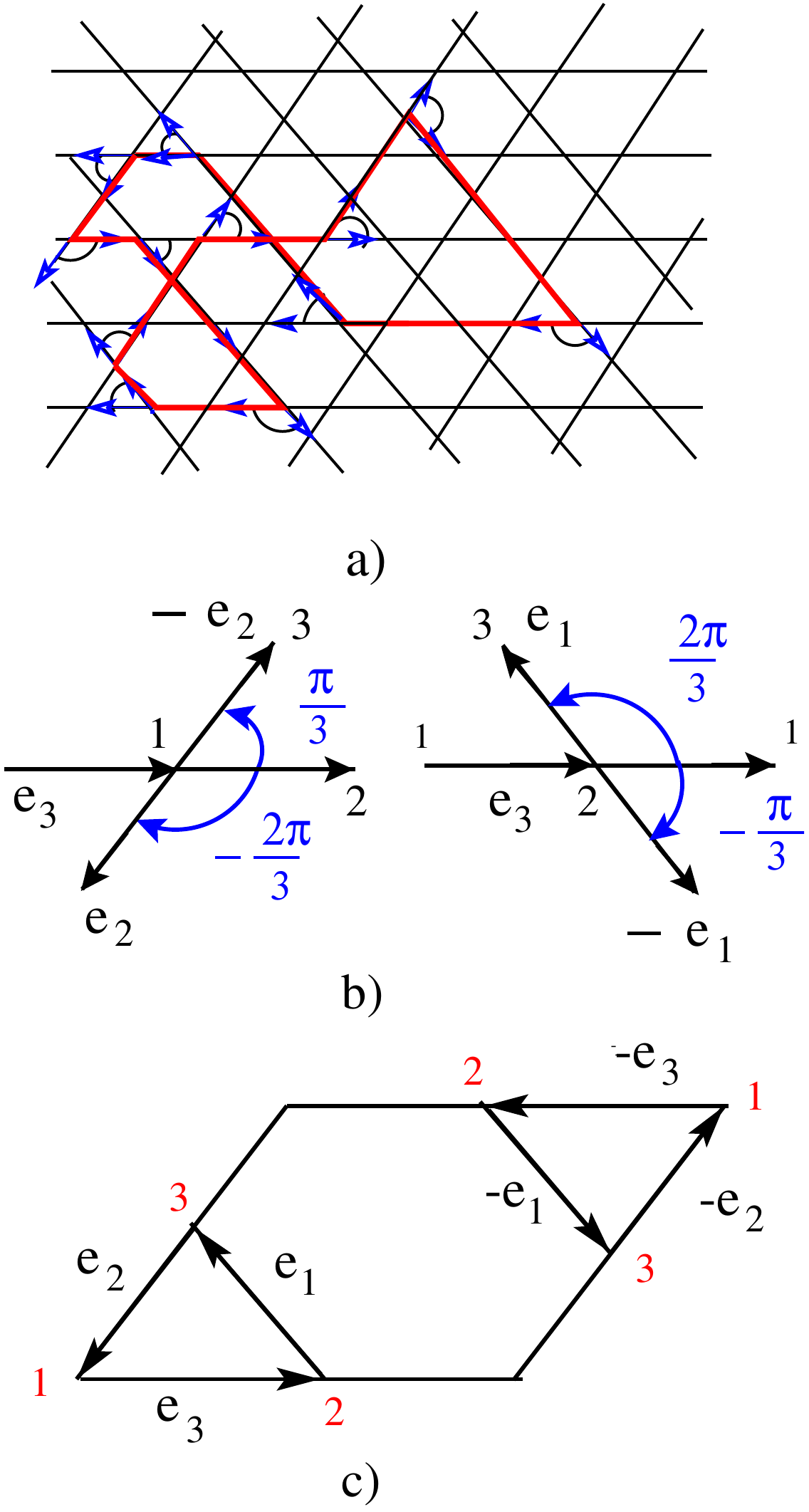}
	\caption{Fermionic trajectories and Kac--Ward rotations on the kagom\'e lattice. (a) An example of an oriented fermionic trajectory on the lattice. (b) Elementary rotations of the fermionic tangent vector at a lattice vertex; the corresponding turning angles are multiples of $\pi/3$. (c) Elementary kagom\'e unit cell containing three inequivalent sites, $1,2,3$, and the three bond directions $\vec e_1,\vec e_2,\vec e_3$, chosen according to the symmetric convention $\vec e_1+\vec e_2+\vec e_3=0$.}
	\label{kagome}
\end{figure}

The explicit hopping part of Eq.~(\ref{S11}) can be written in a
compact form by introducing $\sigma=\pm1$.  According to the
connectivity of Fig.~\ref{kagome}, the $e_3$ links connect
sublattices $1$ and $2$, the $e_1$ links connect sublattices $2$
and $3$, while the $e_2$ links connect sublattices $3$ and $1$.
The complete hopping action in one unit cell of kagom\'e lattice is therefore
\begin{widetext}
\bea
&&H_K^{\rm hop}=\sum_{\vec r}\sum_{\sigma=\pm1} H^{(K)}_{{\vec r},\sigma} ,\nn\\
&&H^{(K)}_{{\vec r},\sigma} =\nn\\
&=&
\lambda_3\,
\bar\psi^{\sigma e_3}_{1,\vec r}
\Big[
\Lambda_{\sigma e_3}^{\sigma e_3}
\psi^{\sigma e_3}_{2,\vec r+\sigma\vec e_3}
+
\Lambda_{\sigma e_3}^{e_1}
\psi^{e_1}_{2,\vec r+\sigma\vec e_3}
+
\Lambda_{\sigma e_3}^{-e_1}
\psi^{-e_1}_{2,\vec r+\sigma\vec e_3}
\Big]
\nn\\
&+&
\lambda_2\,
\bar\psi^{\sigma e_2}_{1,\vec r}
\Big[
\Lambda_{\sigma e_2}^{\sigma e_2}
\psi^{\sigma e_2}_{3,\vec r+\sigma\vec e_2}
+
\Lambda_{\sigma e_2}^{e_1}
\psi^{e_1}_{3,\vec r+\sigma\vec e_2}
+
\Lambda_{\sigma e_2}^{-e_1}
\psi^{-e_1}_{3,\vec r+\sigma\vec e_2}
\Big]
\nn\\
&+&
\lambda_1\,
\bar\psi^{\sigma e_1}_{2,\vec r}
\Big[
\Lambda_{\sigma e_1}^{\sigma e_1}
\psi^{\sigma e_1}_{3,\vec r+\sigma\vec e_1}
+
\Lambda_{\sigma e_1}^{e_2}
\psi^{e_2}_{3,\vec r+\sigma\vec e_1}
+
\Lambda_{\sigma e_1}^{-e_2}
\psi^{-e_2}_{3,\vec r+\sigma\vec e_1}
\Big]
\nn\\
&+&
\lambda_3\,
\bar\psi^{\sigma e_3}_{2,\vec r}
\Big[
\Lambda_{\sigma e_3}^{\sigma e_3}
\psi^{\sigma e_3}_{1,\vec r+\sigma\vec e_3}
+
\Lambda_{\sigma e_3}^{e_2}
\psi^{e_2}_{1,\vec r+\sigma\vec e_3}
+
\Lambda_{\sigma e_3}^{-e_2}
\psi^{-e_2}_{1,\vec r+\sigma\vec e_3}
\Big]
\nn\\
&+&
\lambda_2\,
\bar\psi^{\sigma e_2}_{3,\vec r}
\Big[
\Lambda_{\sigma e_2}^{\sigma e_2}
\psi^{\sigma e_2}_{1,\vec r+\sigma\vec e_2}
+
\Lambda_{\sigma e_2}^{e_3}
\psi^{e_3}_{1,\vec r+\sigma\vec e_2}
+
\Lambda_{\sigma e_2}^{-e_3}
\psi^{-e_3}_{1,\vec r+\sigma\vec e_2}
\Big]
\nn\\
&+&
\lambda_1\,
\bar\psi^{\sigma e_1}_{3,\vec r}
\Big[
\Lambda_{\sigma e_1}^{\sigma e_1}
\psi^{\sigma e_1}_{2,\vec r+\sigma\vec e_1}
+
\Lambda_{\sigma e_1}^{e_3}
\psi^{e_3}_{2,\vec r+\sigma\vec e_1}
+
\Lambda_{\sigma e_1}^{-e_3}
\psi^{-e_3}_{2,\vec r+\sigma\vec e_1}
\Big].
\label{kagome-hop}
\ena
\end{widetext}

The first three terms describe propagation along one orientation
of the elementary links connecting the three sites of the unit
cell, whereas the last three terms correspond to the opposite
orientation. In each term the direction opposite to the incoming
link is absent, in accordance with the no-backtracking condition
of the Kac--Ward construction.

The factors $\Lambda_{e_i}^{e_j}$ describe the rotation of the
fermionic state when the trajectory changes its direction from
$e_i$ to $e_j$. For the {kagom\'e} geometry the elementary angles
between the corresponding directed links are multiples of
$\pi/3$. Therefore, in a fixed eigen-sector of $\sigma_z$, one can
write
\bea
\Lambda_{e_i}^{e_j}
=
\exp\left\{
\frac{i}{2}\phi_{ij}
\right\},
\ena
where $\phi_{ij}$ is the oriented angle between the two successive
links. In particular,
\bea
\Lambda_{e_i}^{e_i}=1,
\qquad
\Lambda_{e_i}^{e_j}
\Lambda_{e_j}^{e_i}=1.
\ena
Products of the $\Lambda$ factors along a closed trajectory give
the KW phase associated with the total rotation of the
tangent vector Fig.5 1a).

As for previous lattices, for the subsequent transformation to momentum space we introduce
$
k_i=\vec k\cdot \vec{e}_i,
\; i=1,2,3.
$
and use convention (\ref{k-relation})
$
k_1+k_2+k_3=0.
$

This symmetric convention will be used below for constructing the
momentum-space fermionic matrix and determining its zero modes and
the critical coupling of the {kagom\'e} Ising model.

%

We now construct the momentum-space fermionic matrix corresponding
to Eq.~(\ref{S11}).  According to Fig.~\ref{kagome}, the links
parallel to $e_3$ connect sublattices $1$ and $2$, the links parallel
to $e_1$ connect sublattices $2$ and $3$, while the links parallel
to $e_2$ connect sublattices $3$ and $1$.

We use the symmetric convention
\begin{equation}
	{\vec e}_1+{\vec e}_2+{\vec e}_3=0,
	\qquad
	k_i=\vec k\cdot {\vec e}_i,
	\qquad
	k_1+k_2+k_3=0.
	\label{kag-krelation}
\end{equation}
The twelve directed fermionic states in one elementary cell are
organized as
\begin{align}
	\Psi_{1,\vec k}
	&=
	\left(
	\psi^{e_2}_{1,\vec k},
	\psi^{-e_2}_{1,\vec k},
	\psi^{e_3}_{1,\vec k},
	\psi^{-e_3}_{1,\vec k}
	\right)^T,
	\nonumber\\
	\Psi_{2,\vec k}
	&=
	\left(
	\psi^{e_3}_{2,\vec k},
	\psi^{-e_3}_{2,\vec k},
	\psi^{e_1}_{2,\vec k},
	\psi^{-e_1}_{2,\vec k}
	\right)^T,
	\nonumber\\
	\Psi_{3,\vec k}
	&=
	\left(
	\psi^{e_1}_{3,\vec k},
	\psi^{-e_1}_{3,\vec k},
	\psi^{e_2}_{3,\vec k},
	\psi^{-e_2}_{3,\vec k}
	\right)^T.
\end{align}
Thus
\begin{equation}
	\Psi_{\vec k}
	=
	\left(
	\Psi_{1,\vec k},
	\Psi_{2,\vec k},
	\Psi_{3,\vec k}
	\right)^T .
	\label{kag-basis}
\end{equation}

We use the Fourier transformation
\begin{equation}
	\psi^{e_i}_{a,\vec r}
	=
	\frac{1}{\sqrt N}
	\sum_{\vec k}
	e^{i\vec k\cdot\vec r}
	\psi^{e_i}_{a,\vec k},
	\qquad
	a=1,2,3 .
\end{equation}
Then the quadratic action takes the form
\begin{equation}
	S_K=
	\sum_{\vec k}
	\bar\Psi_{\vec k}\,
	{\cal A}(\vec k)\,
	\Psi_{\vec k}.
\end{equation}
For the orientation of Fig.~\ref{kagome}, we choose
$	\arg e_3=0,
	\quad
	\arg e_1=\frac{2\pi}{3},
	\quad
	\arg e_2=-\frac{2\pi}{3},
$
and introduce
\begin{equation}
	u=e^{i\pi/6},
	\qquad
	z_i=e^{ik_i}.
	\label{kag-uz}
\end{equation}
The condition (\ref{kag-krelation}) gives
\begin{equation}
	z_1z_2z_3=1.
\end{equation}

The Kac--Ward matrix can conveniently be represented as a
$3\times3$ matrix of $4\times4$ blocks,
\begin{equation}
	{\cal A}(\vec k)
	=
	\begin{pmatrix}
		-I_4&B_{12}&B_{13}\\
		B_{21}&-I_4&B_{23}\\
		B_{31}&B_{32}&-I_4
	\end{pmatrix}.
	\label{kag-A}
\end{equation}
Here the off-diagonal blocks follow directly from the allowed
non-backtracking transitions of Eq.~(\ref{S11}).

The block connecting sublattice $1$ to sublattice $2$ is
\begin{equation}
	B_{12}
	=
	\lambda_3
	\begin{pmatrix}
		0&0&0&0\\
		0&0&0&0\\
		z_3&0&u^2z_3&u^{-1}z_3\\
		0&z_3^{-1}&u^{-1}z_3^{-1}&u^2z_3^{-1}
	\end{pmatrix},
	\label{kag-B12}
\end{equation}
while
\begin{equation}
	B_{21}
	=
	\lambda_3
	\begin{pmatrix}
		u^{-2}z_3&uz_3&z_3&0\\
		uz_3^{-1}&u^{-2}z_3^{-1}&0&z_3^{-1}\\
		0&0&0&0\\
		0&0&0&0
	\end{pmatrix}.
	\label{kag-B21}
\end{equation}

For the links parallel to $e_2$ one obtains
\begin{equation}
	B_{13}
	=
	\lambda_2
	\begin{pmatrix}
		u^{-2}z_2&uz_2&z_2&0\\
		uz_2^{-1}&u^{-2}z_2^{-1}&0&z_2^{-1}\\
		0&0&0&0\\
		0&0&0&0
	\end{pmatrix},
	\label{kag-B13}
\end{equation}
and
\begin{equation}
	B_{31}
	=
	\lambda_2
	\begin{pmatrix}
		0&0&0&0\\
		0&0&0&0\\
		z_2&0&u^2z_2&u^{-1}z_2\\
		0&z_2^{-1}&u^{-1}z_2^{-1}&u^2z_2^{-1}
	\end{pmatrix}.
	\label{kag-B31}
\end{equation}

Finally, the $e_1$ links give
\begin{equation}
	B_{23}
	=
	\lambda_1
	\begin{pmatrix}
		0&0&0&0\\
		0&0&0&0\\
		z_1&0&u^2z_1&u^{-1}z_1\\
		0&z_1^{-1}&u^{-1}z_1^{-1}&u^2z_1^{-1}
	\end{pmatrix},
	\label{kag-B23}
\end{equation}
and
\begin{equation}
	B_{32}
	=
	\lambda_1
	\begin{pmatrix}
		u^{-2}z_1&uz_1&z_1&0\\
		uz_1^{-1}&u^{-2}z_1^{-1}&0&z_1^{-1}\\
		0&0&0&0\\
		0&0&0&0
	\end{pmatrix}.
	\label{kag-B32}
\end{equation}
All vanishing matrix elements which would correspond to an
immediate reversal of a directed link implement the Kac--Ward
no-backtracking condition.

A direct calculation of the determinant of the matrix
(\ref{kag-A}) gives a particularly simple momentum dependence,
\begin{equation}
		\det[{\cal A}(\vec k)]
		=
		{\cal D}_0
		-
		{\cal D}_1\cos(2k_1)
		-
		{\cal D}_2\cos(2k_2)
		-
		{\cal D}_3\cos(2k_3).
	\label{kag-det-general}
\end{equation}
The momentum-independent coefficient is
\begin{align}
	{\cal D}_0
	={}&
	(1+\lambda_1^4)
	(1+\lambda_2^4)
	(1+\lambda_3^4)
	\nonumber\\
	&
	+
	4\lambda_1\lambda_2\lambda_3
	(1+\lambda_1^2)
	(1+\lambda_2^2)
	(1+\lambda_3^2)
	\nonumber\\
	&
	+
	24\lambda_1^2\lambda_2^2\lambda_3^2 .
	\label{kag-D0}
\end{align}
The three momentum-dependent coefficients are
\begin{align}
	{\cal D}_1
	={}&
	2\lambda_1
	(1-\lambda_2^2)
	(1-\lambda_3^2)
	\nonumber\\
	&\times
	\Big[
	\lambda_1
	(1+\lambda_2^2)
	(1+\lambda_3^2)
	+
	2\lambda_2\lambda_3
	(1+\lambda_1^2)
	\Big],
	\label{kag-D1}
	\\[2mm]
	{\cal D}_2
	={}&
	2\lambda_2
	(1-\lambda_1^2)
	(1-\lambda_3^2)
	\nonumber\\
	&\times
	\Big[
	\lambda_2
	(1+\lambda_1^2)
	(1+\lambda_3^2)
	+
	2\lambda_1\lambda_3
	(1+\lambda_2^2)
	\Big],
	\label{kag-D2}
	\\[2mm]
	{\cal D}_3
	={}&
	2\lambda_3
	(1-\lambda_1^2)
	(1-\lambda_2^2)
	\nonumber\\
	&\times
	\Big[
	\lambda_3
	(1+\lambda_1^2)
	(1+\lambda_2^2)
	+
	2\lambda_1\lambda_2
	(1+\lambda_3^2)
	\Big].
	\label{kag-D3}
\end{align}

Therefore the spectral equation of the anisotropic {kagom\'e}
Ising model is
\begin{eqnarray}
&&{\cal D}_0-{\cal D}_1\cos(2k_1)-{\cal D}_2\cos(2k_2)-{\cal D}_3\cos(2k_3)
		=0,\nn\\
&&k_1+k_2+k_3=0.
	\label{kag-spectrum-general}
\end{eqnarray}

The appearance of $2k_i$ rather than $k_i$ has a simple geometrical
origin. Translation from a given sublattice back to the same
sublattice along one of the three straight {kagom\'e} directions
requires two elementary bonds. Consequently, the closed
momentum-dependent contributions contain $z_i^2+z_i^{-2}
=2\cos(2k_i)$.

 It is easy to see, that for $1 >\lambda_i > 0$ all 
${\cal D}_i >0 $
are positive and the spectral equation (\ref{kag-spectrum-general}) will have a first real solution for momenta, when the cosine terms attain their maximum values. This  will happen at 
the center of the Brillouin zone,
\begin{equation}
	k_1=k_2=k_3=0,
\end{equation}
where spectrum will be gapless. At this point
the determinant has the remarkable factorized form
\begin{equation}
		\det{\cal A}(0)
		=
		{\cal Q}^2,
	\label{kag-det-zero}
\end{equation}
where
\begin{align}
	{\cal Q}
	={}&
	1-\lambda_1^2-\lambda_2^2-\lambda_3^2
	-\lambda_1^2\lambda_2^2
	-\lambda_2^2\lambda_3^2
	-\lambda_3^2\lambda_1^2
	\nonumber\\
	&
	+\lambda_1^2\lambda_2^2\lambda_3^2
	-4\lambda_1\lambda_2\lambda_3 .
	\label{kag-Q}
\end{align}
Thus the ferromagnetic critical surface is determined by
\begin{eqnarray}
	\label{kag-critical-lambda}
		1-\lambda_1^2-\lambda_2^2-\lambda_3^2
		&-&\lambda_1^2\lambda_2^2
		-\lambda_2^2\lambda_3^2
		-\lambda_3^2\lambda_1^2 \\
		&+&\lambda_1^2\lambda_2^2\lambda_3^2
		-4\lambda_1\lambda_2\lambda_3
		=0.\nn
\end{eqnarray}
Using
\begin{eqnarray}
\label{kag-1}
	\lambda_i=\tanh J_i,
	\qquad
	\sinh(2J_i)=\frac{2\lambda_i}{1-\lambda_i^2},
\end{eqnarray}
Eq.~(\ref{kag-critical-lambda}) can equivalently be written as
\begin{equation}
		\sum_{i=1}^{3}\cosh(2J_i)
		-
		\prod_{i=1}^{3}\cosh(2J_i)
		-
		\prod_{i=1}^{3}\sinh(2J_i)
		=0.
	\label{kag-critical-J}
\end{equation}
This surface equation for critical coupling was obtained in \cite{Sedrakyan-2026}


For the isotropic model
$\lambda_1=\lambda_2=\lambda_3=\lambda,$
all three coefficients ${\cal D}_i$ become equal. Introducing
\begin{equation}
	C_K(\vec k)
	=
	\cos(2k_1)+\cos(2k_2)+\cos(2k_3),
\end{equation}
the determinant reduces to
\begin{eqnarray}
	\det[{\cal A}(\vec k)]
	&=&(1+\lambda)^4
	\Big[
	(1-\lambda+\lambda^2)^4
	+3\lambda^4
	\nonumber\\
	&-& 2\lambda^2(1-\lambda)^2(1+\lambda^2)
	C_K(\vec k)
	\Big].
	\label{kag-det-isotropic}
\end{eqnarray}
Consequently, the isotropic spectral equation can be written as
\begin{equation}
		\cos(2k_1)+\cos(2k_2)+\cos(2k_3)
		=
		\frac{
			(1-\lambda+\lambda^2)^4+3\lambda^4
		}{
			2\lambda^2(1-\lambda)^2(1+\lambda^2)
		}.
\label{kag-spectrum-isotropic}
\end{equation}
Denoting $k=k_3=-(k_1+k_2)$ as momentum on kagom\'e lattice, and $k_0=(k_1-k_2)/\sqrt{3}$ as the transverse Euclidean momentum
the spectral equation (\ref{kag-spectrum-general}) can be written as
\bea
\label{SE5}
2\cos(\sqrt3 k_0)\cos k+\cos(2k)
=
\frac{(1-\lambda+\lambda^2)^4+3\lambda^4}{2\lambda^2(1-\lambda)^2(1+\lambda^2)}.\qquad
\label{kag-spectrum-k0k}
\ena
For real momenta
\begin{equation}
	C_K(\vec k)\leq 3,
\end{equation}
and the maximum is reached at $\vec k=0$.  At this point
\begin{equation}
	\det{\cal A}(0)
	=
	(1+\lambda)^4
	\left(
	\lambda^4-2\lambda^3-2\lambda+1
	\right)^2.
\end{equation}
The physical ferromagnetic critical coupling therefore satisfies
\begin{equation}
	\lambda_c^4-2\lambda_c^3-2\lambda_c+1=0.
	\label{kag-isotropic-critical}
\end{equation}
The root in the physical interval $0<\lambda_c<1$ is
\begin{equation}
	\label{lambdaK}
		\lambda_c
		=
		\frac{
			1+\sqrt3-\sqrt{2\sqrt3}
		}{2}.
\end{equation}
Finally, since $\lambda_c=\tanh J_c$, one obtains
\begin{equation}
		J_c
		=
		\frac14\ln\left(3+2\sqrt3\right),\; \sinh[2 J_c]=\frac{\sqrt 2}{3^{1/4}}
\end{equation}
This value for critical coupling was obtained earlier in \cite{Syozi-1951,Naya-1954,Matveev-1995}. Expanding Eq.~ (\ref{SE5}) near the critical point $k_0=k=0$ we will get
the spectral equation in its low-energy form
\bea
k_0^2+k^2+m^2(\lambda)=0,
\ena
where
\bea
m^2(\lambda)
	=
	\frac{(\lambda^4-2\lambda^3-2\lambda+1)^2}{6\lambda^2(1-\lambda)^2(1+\lambda^2)}.
\ena
Since
$\lambda=\tanh J,
\quad
\delta\lambda
\simeq
(1-\lambda_c^2)(J-J_c),$
gap becomes especially simple: 
\[m^2\simeq12(J-J_c)^2.\]

In anisotropic case the spectral equation takes the form
\bea
	({\cal D}_1+{\cal D}_2)
	\cos(\sqrt3 k_0)\cos k
	&+&
	({\cal D}_1-{\cal D}_2)
	\sin(\sqrt3 k_0)\sin k \nn\\
	&+&
	{\cal D}_3\cos(2k)
	=
	{\cal D}_0 .
\label{kag-spectrum-anisotropic-k0k}
\ena
In contrast to the isotropic case, the anisotropy generates the
additional term
\bea
({\cal D}_1-{\cal D}_2)
\sin(\sqrt3 k_0)\sin k.
\ena
For $\lambda_1=\lambda_2=\lambda_3$, one has
${\cal D}_1={\cal D}_2={\cal D}_3$, and this term vanishes,
recovering the isotropic spectral equation.

Expanding Eq.~(\ref{kag-spectrum-anisotropic-k0k}) in anisotropic case 
around $k_0=k=0$ up to second order one will obtain
\bea
\label{displ}
	\left(
	k_0-
	\frac{{\cal D}_1-{\cal D}_2}
	{\sqrt3({\cal D}_1+{\cal D}_2)}k
	\right)^2
	+
	v^2k^2+m^2=0,
\ena
 where
\bea
v^2
=
\frac{
	4\left[
	{\cal D}_1{\cal D}_2+
	{\cal D}_3({\cal D}_1+{\cal D}_2)
	\right]
}{
	3({\cal D}_1+{\cal D}_2)^2
},
\ena
and
\bea
m^2
=
\frac{
	2\left[
	{\cal D}_0-{\cal D}_1-{\cal D}_2-{\cal D}_3
	\right]
}{
	3({\cal D}_1+{\cal D}_2)
}.
\ena
which indicates that system is gapless on the critical surface (\ref{kag-critical-J})
defined by ${\cal D}_0-{\cal D}_1-{\cal D}_2-{\cal D}_3=0$.
 The shift in \(k_0\) does not imply that the position of the gap itself is shifted. Rather, the spectrum becomes anisotropic, so that its principal axes do not, in general, coincide with the coordinate axes chosen in the calculation. Accordingly, the correlation lengths become direction dependent and are characterized by different values along the principal axes.

\subsection{Spontaneous magnetization of the 2DIM on the
	kagom\'e lattice}

After obtaining an
exact solution for the free energy in the Ising model on the kagom\'e
lattice, Naya \cite{Naya-1954} made the first attempt to study spontaneous magnetization on this lattice and obtained the first expression for the isotropic case when the exchange interactions in all three directions
in the kagom\'e lattice are the same. 
The generalization of spontaneous magnetization to the anisotropic
case was performed by Syôzi and Naya in \cite{Syozi-1960,Syozi-1960-2}.  Later Syôzi \cite{Syozi-1972}  introduced some amendments to the resulting formula. Recent development 
of the study of spontaneous magnetization of Ising model on kagom\'e
appears in \cite{Kassan-Ogly-2023}.

We calculate the magnetization from the correlation Pfaffian of the
original kagom\'e spins. In the composite basis $\mu=(i,\alpha)$,
$i=1,2,3$ labels the site and $\alpha=1,\ldots,4$ its directed component.
The Green functions are

\begin{align}
G_{\mu\nu}(\mathbf R)
&=\int_{\rm BZ}d\mu(\mathbf k)\,e^{i\mathbf k\cdot\mathbf R}
[\mathcal A^{-1}(\mathbf k)]_{\mu\nu}\nonumber\\
&=\int_{\rm BZ}d\mu(\mathbf k)\,e^{i\mathbf k\cdot\mathbf R}
\frac{\operatorname{Cof}_{\nu\mu}\mathcal A(\mathbf k)}{\mathcal P_K(\mathbf k)}.
\label{kag-G}
\end{align}

For sublattice $i$, the correlator is the full path Pfaffian
$\Gamma_R^{(i)}=\Gamma_\gamma$ of Eq.~(\ref{path-Pf-main}); Appendix~\ref{app:path-pfaffian} relates it to these Green functions:
\bea
\langle s_{i,0}s_{i,R}\rangle
=\operatorname{Pf}\Gamma_R^{(i)},
\qquad
M_i^2=\lim_{R\to\infty}\operatorname{Pf}\Gamma_R^{(i)}.
\label{kag-Pf-def}
\ena
Thus the complete Pfaffian, not one Green-function component or
$\det{\cal A}$ alone, fixes the physical normalization.

The common denominator is
\bea
{\cal P}_K(k_0,k)\equiv\det{\cal A}(k_0,k).
\ena
In the basis $k=k_3=-(k_1+k_2)$ and
$k_0=(k_1-k_2)/\sqrt3$, one has
\bea
{\cal P}_K(k_0,k)
&=&
{\cal D}_0
-({\cal D}_1+{\cal D}_2)\cos(\sqrt3 k_0)\cos k
\nn\\
&&
-({\cal D}_1-{\cal D}_2)\sin(\sqrt3 k_0)\sin k
\nn\\
&&-{\cal D}_3\cos(2k).
\label{kag-P-k0}
\ena
\update{The normalized angular integral is}
\bea
\int\frac{d(\sqrt3 k_0)}{2\pi}
\frac{1}{{\cal P}_K(k_0,k)}
=
\frac{1}{\sqrt{{\cal R}(k)}},
\label{kag-k0-int}
\ena
where
\bea
{\cal R}(k)
&=&
\left[{\cal D}_0-{\cal D}_3\cos(2k)\right]^2
-({\cal D}_1+{\cal D}_2)^2\cos^2 k
\nn\\
&&
-({\cal D}_1-{\cal D}_2)^2\sin^2 k
\nn\\
&=&
{\cal D}_3^2\cos^2(2k)
-2({\cal D}_0{\cal D}_3+{\cal D}_1{\cal D}_2)\cos(2k)\nn\\
&&+{\cal D}_0^2-{\cal D}_1^2-{\cal D}_2^2 .
\label{kag-R}
\ena
Introducing
\bea
X_K={\cal D}_0^2-{\cal D}_1^2-{\cal D}_2^2-{\cal D}_3^2,
\label{kag-X}
\ena
and
\bea
Y_K^2
&=&
({\cal D}_0-{\cal D}_1-{\cal D}_2-{\cal D}_3)\nn\\
&&\times({\cal D}_0-{\cal D}_1+{\cal D}_2+{\cal D}_3)
\nn\\
&&\times
({\cal D}_0+{\cal D}_1-{\cal D}_2+{\cal D}_3)\nn\\
&&\times({\cal D}_0+{\cal D}_1+{\cal D}_2-{\cal D}_3),
\label{kag-Y}
\ena
the elliptic parameter is
\bea
\kappa_K^2=\frac{X_K-Y_K}{X_K+Y_K}.
\label{kag-kappa}
\ena
For $J_i>0$ the ordered region is $\mathcal Q<0$; choosing $Y_K>0$ gives $0\leq\kappa_K^2<1$. The common factor is the effective honeycomb magnetization:
\bea
M_{\rm bulk}
=(1-\kappa_K^2)^{1/8}
=\left[\frac{2Y_K}{X_K+Y_K}\right]^{1/8}.
\label{kag-M-hc}
\ena
The original kagom\'e spin differs by a local insertion retained in the Gaussian elimination.

Using the local Gaussian block of Appendix~\ref{app:local-schur}, let
\bea
\lambda_i=\tanh J_i,
\qquad
d=1+\lambda_1\lambda_2\lambda_3,
\label{kag-local-d}
\ena
and define
\bea
a_1&=&\lambda_1+\lambda_2\lambda_3,\nn\\
a_2&=&\lambda_2+\lambda_1\lambda_3,\nn\\
a_3&=&\lambda_3+\lambda_1\lambda_2.
\label{kag-local-a}
\ena
The Schur complement in Eq.~(\ref{local-triangle-schur}) gives $a_i/d$ and defines positive leg parameters $t_i$ by
\bea
t_2t_3=\frac{a_1}{d},\qquad
t_3t_1=\frac{a_2}{d},\qquad
t_1t_2=\frac{a_3}{d},
\ena
or equivalently
\bea
t_i^2=\frac{a_j a_k}{d\,a_i},
\qquad \{i,j,k\}=\{1,2,3\}.
\label{kag-ti}
\ena
The corresponding odd insertion, Eq.~(\ref{local-decorating-insertion}), has normalized coefficient
\bea
\rho_i=\frac{2t_i}{1+t_i^2}
=\frac{2\sqrt{d\,a_1a_2a_3}}
{d\,a_i+a_j a_k}.
\label{kag-rhoi}
\ena

For two distinct original spins on auxiliary honeycomb edges
$e=(a,b)$ and $f=(c,d)$, the full Pfaffian obeys the exact finite-distance identity
\begin{equation}
\operatorname{Pf}\Gamma^K_{ef}
=\frac{\rho_i\rho_j}{4}
\sum_{p\in\{a,b\}}\sum_{q\in\{c,d\}}
\operatorname{Pf}\Gamma^h_{pq}.
\label{kag-finite-Pf-reduction}
\end{equation}
Each matrix is the path Pfaffian of Eq.~(\ref{path-Pf-main}) on the
corresponding lattice. Appendices~\ref{app:local-schur} and
\ref{app:spectral-link} derive the reduction and identify the same spectral
modulus; hence

\bea
\lim_{R\to\infty}\operatorname{Pf}\Gamma_R^{(i)}
=\rho_i^2\,(1-\kappa_K^2)^{1/4},
\label{kag-Pf-asympt}
\ena
and the three physical sublattice magnetizations are
\begin{align}
M_i&=\rho_i(1-\kappa_K^2)^{1/8}\nonumber\\
&=\frac{2\sqrt{d\,a_1a_2a_3}}{d\,a_i+a_ja_k}
(1-\kappa_K^2)^{1/8},\quad i=1,2,3.
\label{kag-M-general}
\end{align}
Thus $\rho_i$ is generated by the full correlation Pfaffian; the
prefactor-free $(1-\kappa_K^2)^{1/8}$ is only the common bulk factor.

If a single magnetization per kagom\'e site is desired, the natural
unit-cell average is
\bea
M_K^{\rm av}=\frac{M_1+M_2+M_3}{3}.
\label{kag-M-average}
\ena
On the critical surface ${\cal D}_0-{\cal D}_1-{\cal D}_2-{\cal D}_3={\cal Q}^2=0$,
so $Y_K=M_{\rm bulk}=0$. Since $\rho_i$ remains finite, all three $M_i$
vanish with $\beta=1/8$.

As an independent check, the exact star--triangle transformation gives
\bea
&&
\exp\!\left(J_1s_2s_3+J_2s_3s_1+J_3s_1s_2\right)
\nn\\
&&\qquad
={\cal A}_\star
\sum_{\tau=\pm1}
\exp\!\left[\tau(L_1s_1+L_2s_2+L_3s_3)\right].
\label{kag-star-triangle}
\ena
Matching the local Gaussian weights gives simply
\bea
t_i=\tanh L_i.
\label{kag-ti-star}
\ena
Equation~(\ref{kag-rhoi}) therefore becomes
\bea
\rho_i=\frac{2\tanh L_i}{1+\tanh^2L_i}
=\tanh(2L_i).
\label{kag-rho-star}
\ena
Thus the star--triangle representation reproduces the same result,
\bea
M_i=\tanh(2L_i)(1-\kappa_K^2)^{1/8},
\ena
with the same local coefficient $\rho_i$ as the Pfaffian reduction.

Isotropically, the prefactor-free expression of Ref.~\cite{Kassan-Ogly-2023} is $M_{\rm bulk}$; the physical spin magnetization also contains $\rho$.

\subsection{Isotropic kagom\'e lattice.}
For
\bea
\lambda_1=\lambda_2=\lambda_3=\lambda,
\qquad \lambda=\tanh J,
\ena
one has
\bea
{\cal D}_0&=&(1+\lambda)^4
\left[(1-\lambda+\lambda^2)^4+3\lambda^4\right],
\\
{\cal D}_1&=&{\cal D}_2={\cal D}_3
=2\lambda^2(1+\lambda)^4(1-\lambda)^2(1+\lambda^2).\qquad\nn
\ena
The common factor $(1+\lambda)^4$ in each $\mathcal D_i$ generates a common factor
$(1+\lambda)^8$ in both $X_K$ and $Y_K$; this factor cancels only from
$\kappa_K^2$. In the ordered phase the physical branch is
\bea
Y_K
&=&
-(1+\lambda)^8
\left(\lambda^4-2\lambda^3-2\lambda+1\right)
\nn\\
&&\times
\left(\lambda^4-2\lambda^3+4\lambda^2-2\lambda+1\right)^3,
\label{kag-Y-physical}
\ena
and
\bea
X_K+Y_K
=
32(1+\lambda)^8\lambda^6(1-\lambda+\lambda^2)^2.
\label{kag-XY-iso}
\ena
Hence the common bulk factor is
\bea
M_{\rm bulk}(\lambda)
=\left[1-\kappa_K^2(\lambda)\right]^{1/8},
\label{kag-M-onsager}
\ena
where
\begin{align}
\kappa_K^2(\lambda)
&=\frac{(1-\lambda)^6(1+\lambda^2)^3}
{16\lambda^6(1-\lambda+\lambda^2)^2}\nonumber\\
&\quad\times(\lambda^4-2\lambda^3+6\lambda^2-2\lambda+1).
\label{kag-kappa-iso}
\end{align}
The local Pfaffian quantities simplify to
\bea
d&=&1+\lambda^3=(1+\lambda)(1-\lambda+\lambda^2),\nn\\
a_1&=&a_2=a_3=\lambda(1+\lambda),
\ena
so that
\bea
t^2=\frac{\lambda}{1-\lambda+\lambda^2},
\ena
and Eq.~(\ref{kag-rhoi}) gives directly
\bea
\rho(\lambda)
=\frac{2\sqrt{\lambda(1-\lambda+\lambda^2)}}{1+\lambda^2}.
\label{kag-rho-iso-lambda}
\ena
Thus the isotropic physical kagom\'e magnetization obtained from the
full fermionic correlation Pfaffian is
\bea
M_K^{\rm iso}
=
\frac{2\sqrt{\lambda(1-\lambda+\lambda^2)}}{1+\lambda^2}
\left[1-\kappa_K^2(\lambda)\right]^{1/8}.
\label{kag-M-iso-lambda}
\ena
Introducing $u=e^{-2J}$, $\lambda=(1-u)/(1+u)$, the Pfaffian amplitude
becomes
\bea
\rho
=
\frac{\sqrt{(1+3u^2)(1-u^2)}}{1+u^2}.
\label{kag-naya-prefactor}
\ena
Consequently
\bea
M_K^{\rm iso}
&=&
\frac{\sqrt{(1+3u^2)(1-u^2)}}{1+u^2}\nn\\
&\times&\left[
1-
\frac{
128u^6(1+u^2)^3(1+3u^4)
}{
(1-u^2)^6(1+3u^2)^2
}
\right]^{1/8}.
\label{kag-M-iso-corrected}
\ena
This is the classical Naya expression~\cite{Naya-1954}.  Its
low-temperature expansion is
\bea
M_K^{\rm iso}=1-2u^4-12u^6+O(u^8),
\label{kag-lowT-series}
\ena
whereas the common bulk factor alone starts as
\bea
M_{\rm bulk}=1-16u^6-48u^8+O(u^{10}).
\ena
The difference is the local Pfaffian amplitude $\rho$. The isotropic
star--triangle relation gives $t=\tanh L$ and $\rho=\tanh(2L)$, confirming
Eq.~(\ref{kag-naya-prefactor}). As $T\to0$, $\rho\to1$; at criticality
$M_K^{\rm iso}$ vanishes with exponent $1/8$.

	%
	%

\section{Critical Surface and Spontaneous Magnetization of the Fully Anisotropic Ising Model on the dice Lattice}
{The planar dual of the kagom\'e lattice is the dice lattice (also known as the $T_3$ lattice or rhombille tiling).}

We now construct the fermionic action for the Ising model on the dice
lattice. The elementary unit cell contains three
inequivalent sites. One of them, denoted by $1$, has coordination
number six, while the other two sites, denoted by $2$ and $3$, have
coordination number three.

We introduce three elementary vectors $\vec e_i$, $i=1,2,3$, chosen
according to the symmetric convention
\bea
\label{dice-vectors}
\vec e_1+\vec e_2+\vec e_3=0.
\ena
The six directed links attached to the central site $1$ are
$\pm\vec e_1,\pm\vec e_2,\pm\vec e_3$. We choose the orientations
of the two three-coordinated sublattices such that the directed
fermionic states associated with the three sites of the unit cell are
\bea
\label{dice-states}
1:\;&\{\pm e_1,\pm e_2,\pm e_3\},\nn\\
2:\;&\{-e_1,-e_2,-e_3\},\nn\\
3:\;&\{e_1,e_2,e_3\}.
\ena
Thus, the elementary unit cell contains altogether
\bea
6+3+3=12
\ena
directed fermionic states.

The connectivity of the lattice can be represented in the form
\bea
\label{dice-connectivity}
1\stackrel{+e_i}{\longrightarrow}2,
\qquad
2\stackrel{-e_i}{\longrightarrow}1,
\nn\\
1\stackrel{-e_i}{\longrightarrow}3,
\qquad
3\stackrel{+e_i}{\longrightarrow}1,
\qquad i=1,2,3.
\ena
Let $J_i$ be the Ising coupling constants associated with the three
inequivalent bond directions and introduce $\lambda_i=\tanh [J_i], \; i=1,2,3.$.
%

\begin{figure}[t]
	\centering
	\includegraphics[width=0.9\linewidth]{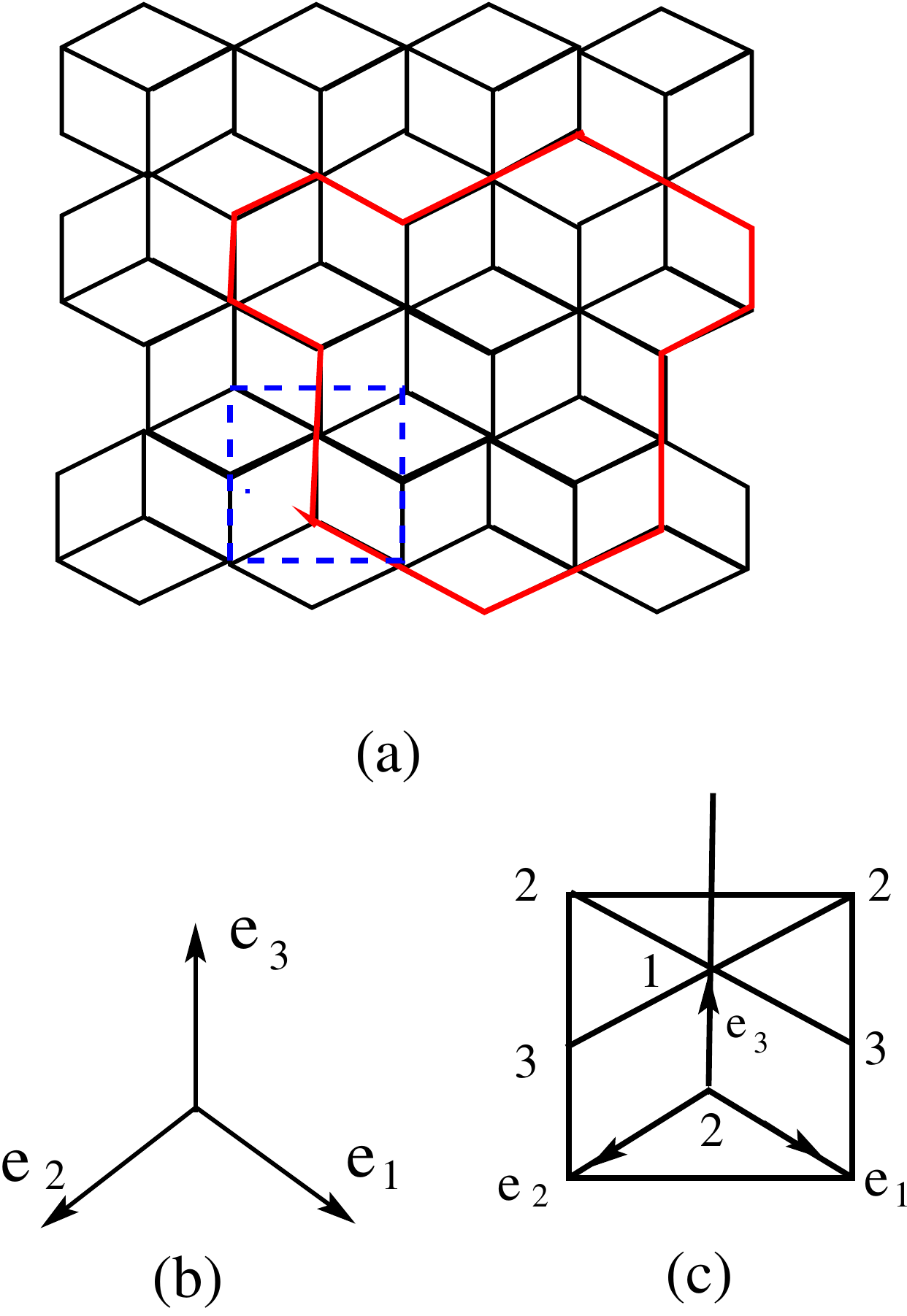}
	\caption{Duality between the kagom\'e lattice and its dual dice ($T_3$) lattice. The vertices of the dual lattice are located at the centers of the triangular and hexagonal plaquettes of the kagom\'e lattice, and each dual link crosses one kagom\'e bond. The dual sites corresponding to triangular plaquettes have coordination number three, whereas those corresponding to hexagonal plaquettes have coordination number six.}
	\label{dual-kagome}
\end{figure}

The hopping part of the action, which, in analogy with (\ref{S1}) and (\ref{S11}) will reproduce sign-factor on dice lattice can
be written in the following compact form
\bea
\label{dice-hop-compact}
H_D^{\rm hop}
&=&
\sum_{\vec r}\sum_{i=1}^{3}\sum_{\sigma=\pm1}
\lambda_i
\Bigg\{
\bar\psi^{\sigma e_i}_{1,\vec r}
\sum_{\substack{j=1\\j\neq i}}^{3}
\Lambda_{\sigma e_i}^{-\sigma e_j}
\psi^{-\sigma e_j}_{a_\sigma,\vec r+\sigma\vec e_i}
\nn\\
&+&
\bar\psi^{-\sigma e_i}_{a_\sigma,\vec r}
\sum_{\substack{j=1,2,3;\;\tau=\pm1\\
		(\tau,j)\neq(\sigma,i)}}
\Lambda_{-\sigma e_i}^{\tau e_j}
\psi^{\tau e_j}_{1,\vec r-\sigma\vec e_i}
\Bigg\}.
\ena
Here $a_{+}=2$ and $a_{-}=3$. The first term describes propagation
from the six-coordinated site $1$ to one of the three-coordinated
sites. Since immediate backtracking is forbidden, only two outgoing
directions are allowed. The second term describes propagation from a
three-coordinated site to the six-coordinated site, where five
outgoing directions are allowed; the condition
$(\tau,j)\neq(\sigma,i)$ excludes the backtracking direction.

The complete fermionic action is therefore
\bea
\label{dice-action-compact}
S_D(\psi)
&=&
H_D^{\rm hop}
-
\sum_{\vec r}\sum_{i=1}^{3}
\Big[
\bar\psi^{e_i}_{1,\vec r}\psi^{e_i}_{1,\vec r}
+
\bar\psi^{-e_i}_{1,\vec r}\psi^{-e_i}_{1,\vec r}
\nn\\
&&\qquad\qquad+
\bar\psi^{-e_i}_{2,\vec r}\psi^{-e_i}_{2,\vec r}
+
\bar\psi^{e_i}_{3,\vec r}\psi^{e_i}_{3,\vec r}
\Big].
\ena

As before, the KW factor is defined by
\bea
\label{dice-KW}
\Lambda_e^{e'}
=
\exp\left[
\frac{i}{2}\phi(e,e')
\right],
\qquad e'\neq -e,
\ena
where $\phi(e,e')$ is the oriented turning angle between two
successive directed links. The condition $e'\neq-e$ excludes
immediate backtracking.


The explicit hopping part of the action, analogous to
Eq.~(\ref{kagome-hop}) for the kagom\'e lattice, can be written as
\bea
\label{dice-hop}
H_D^{\rm hop}
=
\sum_{\vec r}\sum_{i=1}^{3}
\left[
{\cal H}^{(12)}_{i,\vec r}
+
{\cal H}^{(13)}_{i,\vec r}
+
{\cal H}^{(21)}_{i,\vec r}
+
{\cal H}^{(31)}_{i,\vec r}
\right].\qquad
\ena
For a fermion propagating from the six-coordinated site $1$ to the
three-coordinated site $2$, one has
\bea
\label{dice-hop-12}
{\cal H}^{(12)}_{i,\vec r}
=
\lambda_i\,
\bar\psi^{e_i}_{1,\vec r}
\sum_{\substack{j=1\\j\neq i}}^{3}
\Lambda_{e_i}^{-e_j}
\psi^{-e_j}_{2,\vec r+\vec e_i}.
\ena
There are only two allowed continuations at the site $2$, since the
direction $-e_i$ would correspond to immediate backtracking.

Similarly, propagation from the central site $1$ to sublattice $3$
is described by
\bea
\label{dice-hop-13}
{\cal H}^{(13)}_{i,\vec r}
=
\lambda_i\,
\bar\psi^{-e_i}_{1,\vec r}
\sum_{\substack{j=1\\j\neq i}}^{3}
\Lambda_{-e_i}^{e_j}
\psi^{e_j}_{3,\vec r-\vec e_i}.
\ena
For propagation from the three-coordinated site $2$ to the
six-coordinated site $1$, five outgoing directions are allowed.
Thus,
\bea
\label{dice-hop-21}
{\cal H}^{(21)}_{i,\vec r}
&=&
\lambda_i\,
\bar\psi^{-e_i}_{2,\vec r}
\Bigg[
\Lambda_{-e_i}^{-e_i}
\psi^{-e_i}_{1,\vec r-\vec e_i}
\\
&&+
\sum_{\substack{j=1\\j\neq i}}^{3}
\left(
\Lambda_{-e_i}^{e_j}
\psi^{e_j}_{1,\vec r-\vec e_i}
+
\Lambda_{-e_i}^{-e_j}
\psi^{-e_j}_{1,\vec r-\vec e_i}
\right)
\Bigg].\nn
\ena
The direction $e_i$ is absent because it corresponds to immediate
backtracking.

Analogously, propagation from sublattice $3$ to the central site
$1$ is
\bea
\label{dice-hop-31}
{\cal H}^{(31)}_{i,\vec r}
&=&
\lambda_i\,
\bar\psi^{e_i}_{3,\vec r}
\Bigg[
\Lambda_{e_i}^{e_i}
\psi^{e_i}_{1,\vec r+\vec e_i}
\\
&&+
\sum_{\substack{j=1\\j\neq i}}^{3}
\left(
\Lambda_{e_i}^{e_j}
\psi^{e_j}_{1,\vec r+\vec e_i}
+
\Lambda_{e_i}^{-e_j}
\psi^{-e_j}_{1,\vec r+\vec e_i}
\right)
\Bigg].\nn
\ena
Here the direction $-e_i$ is excluded by the no-backtracking
condition.

A fermion arriving at a three-coordinated site has two allowed
continuations, whereas a fermion arriving at the six-coordinated
site has five allowed continuations. This structure naturally leads
to a $12\times12$ fermionic matrix in momentum space.

A convenient ordering of the twelve fermionic fields is
\bea
\label{dice-basis}
\Psi_{\vec r}
=
\Big(
&&\psi^{e_1}_{1,\vec r},
\psi^{-e_1}_{1,\vec r},
\psi^{e_2}_{1,\vec r},
\psi^{-e_2}_{1,\vec r},
\psi^{e_3}_{1,\vec r},
\psi^{-e_3}_{1,\vec r},
\nn\\
&&
\psi^{-e_1}_{2,\vec r},
\psi^{-e_2}_{2,\vec r},
\psi^{-e_3}_{2,\vec r},
\psi^{e_1}_{3,\vec r},
\psi^{e_2}_{3,\vec r},
\psi^{e_3}_{3,\vec r}
\Big)^T.
\ena
With the Fourier transformation
\bea
\psi^{e}_{a,\vec r}
=
\frac{1}{\sqrt{N}}
\sum_{\vec k}
e^{i\vec k\cdot\vec r}
\psi^{e}_{a,\vec k},
\ena
the action takes the quadratic momentum-space form
\bea
S_D(\psi)
=
\sum_{\vec k}
\bar\Psi_{\vec k}\,
{\cal A}_D(\vec k)\,
\Psi_{\vec k},
\ena
where ${\cal A}_D(\vec k)$ is a $12\times12$ matrix.

We introduce
\bea
t=e^{i\pi/6},
\quad
z_i=e^{ik_i},
\quad
k_1+k_2+k_3=0.
\ena
In the basis
\bea
\Psi_{\bf k}
=
\Big(
&&\psi^{e_1}_{1,\bf k},
\psi^{-e_1}_{1,\bf k},
\psi^{e_2}_{1,\bf k},
\psi^{-e_2}_{1,\bf k},
\psi^{e_3}_{1,\bf k},
\psi^{-e_3}_{1,\bf k},
\nn\\
&&
\psi^{-e_1}_{2,\bf k},
\psi^{-e_2}_{2,\bf k},
\psi^{-e_3}_{2,\bf k},
\psi^{e_1}_{3,\bf k},
\psi^{e_2}_{3,\bf k},
\psi^{e_3}_{3,\bf k}
\Big)^T ,
\label{dice-momentum-basis}
\ena
the momentum-space action is
\bea
S_D
=
\sum_{\bf k}
\bar\Psi_{\bf k}
{\cal A}_D({\bf k})
\Psi_{\bf k},
\ena
where
\bea
{\cal A}_D({\bf k})
=
\left(
\begin{array}{ccc}
	-I_6&B_{12}&B_{13}\\
	B_{21}&-I_3&0\\
	B_{31}&0&-I_3
\end{array}
\right).
\label{dice-A-block}
\ena
The corresponding blocks are
\bea
B_{12}
=
\left(
\begin{array}{ccc}
	0&\lambda_1t^{-1}z_1&\lambda_1tz_1\\
	0&0&0\\
	\lambda_2tz_2&0&\lambda_2t^{-1}z_2\\
	0&0&0\\
	\lambda_3t^{-1}z_3&\lambda_3tz_3&0\\
	0&0&0
\end{array}
\right),
\ena
\bea
B_{13}
=
\left(
\begin{array}{ccc}
	0&0&0\\
	0&\lambda_1t^{-1}z_1^{-1}&\lambda_1tz_1^{-1}\\
	0&0&0\\
	\lambda_2tz_2^{-1}&0&\lambda_2t^{-1}z_2^{-1}\\
	0&0&0\\
	\lambda_3t^{-1}z_3^{-1}&\lambda_3tz_3^{-1}&0
\end{array}
\right),
\ena
\bea
B_{21}
=
\left(
\begin{array}{cccccc}
	0&
	\frac{\lambda_1}{z_1}&
	\frac{\lambda_1}{t z_1}&
	\frac{\lambda_1 t^2}{z_1}&
	\frac{\lambda_1 t}{z_1}&
	\frac{\lambda_1}{t^2 z_1}
	\\
	\frac{\lambda_2 t}{z_2}&
	\frac{\lambda_2}{t^2 z_2}&
	0&
	\frac{\lambda_2}{z_2}&
	\frac{\lambda_2}{t z_2}&
	\frac{\lambda_2 t^2}{z_2}
	\\
	\frac{\lambda_3}{t z_3}&
	\frac{\lambda_3 t^2}{z_3}&
	\frac{\lambda_3 t}{z_3}&
	\frac{\lambda_3}{t^{2}z_3}&
	0&
	\frac{\lambda_3}{z_3}
\end{array}
\right),
\ena

and
\bea
\label{31}
B_{31}
=
\left(
\begin{array}{cccccc}
	\lambda_1 z_1&
	0&
	\lambda_1 t^2 z_1&
	\frac{\lambda_1 z_1}{t}&
	\frac{\lambda_1 z_1}{t^2}&
	\lambda_1 t z_1
	\\
	\frac{\lambda_2 z_2}{t^2}&
	\lambda_2 t z_2&
	\lambda_2 z_2&
	0&
	\lambda_2 t^2 z_2&
	\frac{\lambda_2 z_2}{t}
	\\
	\lambda_3 t^2 z_3&
	\frac{\lambda_3 z_3}{t}&
	\frac{\lambda_3 z_3}{t^2 }&
	\lambda_3 t z_3&
	\lambda_3 z_3&
	0
\end{array}
\right).\nn\\
\ena
The full $12 \times 12$  size matrix reads 
\begin{widetext}
\begin{equation}
\label{dice-A-full}
\resizebox{0.98\textwidth}{!}{$
{\cal A}_D({\bf k})=
\left(
		\begin{array}{cccccccccccc}
			-1&0&0&0&0&0&
			0&\lambda_1t^{-1}z_1&\lambda_1tz_1&
			0 & 0 & 0
			\\
			0&-1&0&0&0&0&
			0&0&0&
			0&\lambda_1t^{-1}z_1^{-1}&\lambda_1tz_1^{-1}
			\\
			0&0&-1&0&0&0&
			\lambda_2tz_2&0&\lambda_2t^{-1}z_2&
			0&0&0
			\\
			0&0&0&-1&0&0&
			0&0&0&
			\lambda_2tz_2^{-1}&0&\lambda_2t^{-1}z_2^{-1}
			\\
			0&0&0&0&-1&0&
			\lambda_3t^{-1}z_3&\lambda_3tz_3&0&
			0&0&0
			\\
			0&0&0&0&0&-1&
			0&0&0&
			\lambda_3t^{-1}z_3^{-1}&\lambda_3tz_3^{-1}&0
			\\
			0&\lambda_1z_1^{-1}&\lambda_1t^{-1}z_1^{-1}&
			\lambda_1t^2z_1^{-1}&\lambda_1tz_1^{-1}&
			\lambda_1t^{-2}z_1^{-1}&
			-1&0&0&0&0&0
			\\
			\lambda_2tz_2^{-1}&\lambda_2t^{-2}z_2^{-1}&0&
			\lambda_2z_2^{-1}&\lambda_2t^{-1}z_2^{-1}&
			\lambda_2t^2z_2^{-1}&
			0&-1&0&0&0&0
			\\
			\lambda_3t^{-1}z_3^{-1}&\lambda_3t^2z_3^{-1}&
			\lambda_3tz_3^{-1}&\lambda_3t^{-2}z_3^{-1}&
			0&\lambda_3z_3^{-1}&
			0&0&-1&0&0&0
			\\
			\lambda_1z_1&0&\lambda_1t^2z_1&
			\lambda_1t^{-1}z_1&\lambda_1t^{-2}z_1&
			\lambda_1tz_1&
			0&0&0&-1&0&0
			\\
			\lambda_2t^{-2}z_2&\lambda_2tz_2&
			\lambda_2z_2&0&
			\lambda_2t^2z_2&\lambda_2t^{-1}z_2&
			0&0&0&0&-1&0
			\\
			\lambda_3t^2z_3&\lambda_3t^{-1}z_3&
			\lambda_3t^{-2}z_3&\lambda_3tz_3&
			\lambda_3z_3&0&
			0&0&0&0&0&-1
		\end{array}
		\right)
$}
\end{equation}
\end{widetext}
The determinant of the $12\times12$ fermionic matrix can be written as
\bea
\label{dice-det}
\det\cal A_D({\bf k})
&=&
{\cal F}_0
-
{\cal F}_{12}\cos(k_1-k_2)
-
{\cal F}_{13}\cos(k_1-k_3)\nn\\
&-&
{\cal F}_{23}\cos(k_2-k_3),
\ena
where
\bea
{\cal F}_0
&=&
1+
6\left(
\lambda_1^2\lambda_2^2+
\lambda_1^2\lambda_3^2+
\lambda_2^2\lambda_3^2
\right)
\nn\\
&&+
\lambda_1^4\lambda_2^4+
\lambda_1^4\lambda_3^4+
\lambda_2^4\lambda_3^4
+
24\lambda_1^2\lambda_2^2\lambda_3^2
\nn\\
&&+
6\lambda_1^2\lambda_2^2\lambda_3^2
\left(
\lambda_1^2+\lambda_2^2+\lambda_3^2
\right),
\ena
and
\bea
{\cal F}_{12}
&=&
4\lambda_1\lambda_2(1-\lambda_3^2)
\left[
1+\lambda_1^2\lambda_2^2
-\lambda_1^2\lambda_3^2
-\lambda_2^2\lambda_3^2
\right],
\nn\\
{\cal F}_{13}
&=&
4\lambda_1\lambda_3(1-\lambda_2^2)
\left[
1+\lambda_1^2\lambda_3^2
-\lambda_1^2\lambda_2^2
-\lambda_2^2\lambda_3^2
\right],
\nn\\
{\cal F}_{23}
&=&
4\lambda_2\lambda_3(1-\lambda_1^2)
\left[
1+\lambda_2^2\lambda_3^2
-\lambda_1^2\lambda_2^2
-\lambda_1^2\lambda_3^2
\right].
\ena
At zero momentum the determinant factorizes as
\bea
\label{dice-det-zero}
&&\det\cal A_D(\rm 0)={\cal Q}^2\nn\\
&&{\cal Q}=
1+
\lambda_1^2\lambda_2^2+
\lambda_1^2\lambda_3^2+
\lambda_2^2\lambda_3^2 
-2\big(\lambda_1\lambda_2\\
&+&
\lambda_1\lambda_3+
\lambda_2\lambda_3\big)
-2\lambda_1\lambda_2\lambda_3
\big(\lambda_1+\lambda_2+\lambda_3\big).\nn 
\ena
Thus, the critical surface of the dice lattice is determined by
\bea
\label{Q}
{\cal Q}=0,
\ena
which in J-terms reads
\bea
\label{dice-critical-J}
&&2-\cosh\left[2(J_1+J_2+J_3)\right]
+\cosh\left[2(J_1+J_2-J_3)\right]
\nn\\
&+&\cosh\left[2(J_1-J_2+J_3)\right]
\nn\\
&+&\cosh\left[2(-J_1+J_2+J_3)\right]
=0.
\ena
This is the critical-coupling equation for the anisotropic dice lattice.
It is dual to the kagom\'e critical-coupling equation~(\ref{kag-critical-J}).
Substituting $e^{-2\bar J}=\tanh J$ into Eq.~(\ref{kag-critical-J}) gives
Eq.~(\ref{dice-critical-J}). In the isotropic case its solution is
\bea
\label{crit}
\cosh[2 J_c^{dice}]=\frac{1+\sqrt{3}}{2},\;\; \sinh[2 J_c^{dice}]=\frac{3^{1/4}}{\sqrt{2}},
\ena
the critical coupling of the Ising model on the dice lattice.

We next derive the low-energy spectral equation for the anisotropic dice lattice. 
Introducing the two Euclidean momentum variables
\bea
\label{diceco}
k=k_3,\qquad k_0=(k_1-k_2)/\sqrt{3},
\ena
and using $k_1+k_2+k_3=0$,  the spectral equation
$\det{\cal A}_D({\bf k})=0$ takes the form
\bea
\label{dice-spectrum}
{\cal F}_{12}\cos \big[\sqrt{3}k_0\big]&+&
{\cal F}_{13}\cos\big[\frac{\sqrt{3}k_0-3k}{2}\big]\nn\\
&+&
{\cal F}_{23}\cos\big[\frac{\sqrt{3}k_0+3k}{2}\big]
={\cal F}_0.\qquad 
\ena
Expanding this expression up to second order around
$k_0=k=0$, we obtain
{
\bea
0&=&
\Delta_D
+\frac{9}{8}({\cal F}_{13}+{\cal F}_{23})k^2
+\frac{3\sqrt{3}}{4}({\cal F}_{23}-{\cal F}_{13})k_0k
\nn\\
&&+
\frac{3}{8}
(4{\cal F}_{12}+{\cal F}_{13}+{\cal F}_{23})k_0^2 ,
\label{dice-spectrum-expansion}
\ena
where
\bea
\Delta_D=
{\cal F}_0-
{\cal F}_{12}-
{\cal F}_{13}-
{\cal F}_{23}.
\ena
For later use it is convenient to express the same coefficients in
hyperbolic variables.  We define
\bea
C_i&=&\cosh(2J_i),\qquad
S_i=\sinh(2J_i),\qquad i=1,2,3,\nn\\
{\cal N}&=&
\frac{8}{(1+C_1)^2(1+C_2)^2(1+C_3)^2},
\ena
for which
\bea
{\cal F}_0
&=&{\cal N}\left[
3+4C_1C_2C_3+\prod_{i=1}^{3}\cosh(4J_i)
\right],
\nn\\
{\cal F}_{12}
&=&4{\cal N}S_1S_2(C_1C_2+C_3),
\nn\\
{\cal F}_{13}
&=&4{\cal N}S_1S_3(C_1C_3+C_2),
\nn\\
{\cal F}_{23}
&=&4{\cal N}S_2S_3(C_2C_3+C_1).
\label{Aij}
\ena
The factor $4$ in ${\cal F}_{12}$, ${\cal F}_{13}$, and
${\cal F}_{23}$ is essential.  With these definitions the hyperbolic
parametrization reproduces the coefficients of the determinant
\update{(\ref{dice-det}) exactly.}  Hence
$\Delta_D=0$ is exactly equivalent to the critical condition
(\ref{Q}).

On the critical surface, Eq.~(\ref{dice-spectrum-expansion}) reduces to
%
\bea
&&3({\cal F}_{13}+{\cal F}_{23})k^2
+2\sqrt{3}({\cal F}_{23}-{\cal F}_{13})k_0k\nn\\
&&+(4{\cal F}_{12}+{\cal F}_{13}+{\cal F}_{23})k_0^2=0.
\ena
For the isotropic dice lattice,
\bea
J_1=J_2=J_3=J,
\;
C=\cosh(2J),
\;
S=\sinh(2J),\qquad
\ena
all three coefficients become equal,
\bea
{\cal F}_{12}={\cal F}_{13}={\cal F}_{23}\equiv {\cal F},
\quad
{\cal F}=4 {\cal N} S^2C(C+1),\qquad
\ena
while
{
\bea
{\cal F}_0
=
{\cal N}\left[3+4C^3+(2C^2-1)^3\right].
\ena
}
The spectral equation therefore reduces to
{
\bea
\cos[\sqrt{3} k_0]
+
2\cos\big[\frac{\sqrt{3}k_0}{2}\big]\cos\big[\frac{3k}{2}\big]
=
\frac{{\cal F}_0}{\cal F}.\qquad\nn\qquad
\ena
}
Expanding around $k_0=k=0$, one finds the low-momentum spectral 
equation as
\bea
k_0^2+ k^2+m^2=0,
\ena
with
\bea
m^2=\frac{2(2C^2-2C-1)^2}{9C(C-1)}.
\ena
At the critical point
\bea
\label{cc}
C_c=\cosh(2J_c)=\frac{1+\sqrt3}{2},
\ena
and $m=0$.

\subsection{Spontaneous magnetization of the Ising model on the
	dice lattice}

We calculate the inequivalent dice-lattice magnetizations from the full
fermionic correlation Pfaffians. The unit cell contains one
six-coordinated site ($1$) and two three-coordinated sites ($2,3$):
\begin{align}
M_a^2&=\lim_{R\to\infty}\langle s_{a,0}s_{a,R}\rangle\nonumber\\
&=\lim_{R\to\infty}\operatorname{Pf}\Gamma_R^{(a)},\qquad a=1,2,3.
\label{dice-M-def}
\end{align}
where $\Gamma_R^{(a)}$ is the path matrix of Eq.~(\ref{path-Pf-main}); Appendix~\ref{app:path-pfaffian} relates it to the $12\times12$ dice Green function. In the basis of Eq.~(\ref{dice-momentum-basis}),

\begin{align}
G_{\mu\nu}(\mathbf R)
&=\int_{\rm BZ}d\mu(\mathbf k)\,e^{i\mathbf k\cdot\mathbf R}
[\mathcal A_D^{-1}(\mathbf k)]_{\mu\nu}\nonumber\\
&=\int_{\rm BZ}d\mu(\mathbf k)\,e^{i\mathbf k\cdot\mathbf R}
\frac{\operatorname{Cof}_{\nu\mu}\mathcal A_D(\mathbf k)}{\mathcal P_D(\mathbf k)}.
\label{dice-green}
\end{align}

with the composite indices $\mu=(a,\alpha)$ and $\nu=(b,\beta)$.
Species $1$ has six directed components and species $2,3$ have three each.

Using Eq.~(\ref{diceco}), the common spectral polynomial can be written,
with the normalization of Eq.~(\ref{Aij}), as
\bea
{\cal P}_D(k_0,k)
&=&
{\cal F}_0
-{\cal F}_{12}\cos(\sqrt{3}k_0)
\nn\\
&&-{\cal F}_{13}
\cos\left(\frac{\sqrt{3}k_0-3k}{2}\right)
\nn\\
&&
-{\cal F}_{23}
\cos\left(\frac{\sqrt{3}k_0+3k}{2}\right),
\label{dice-P}
\ena
where ${\cal F}_{ij}$ are the full coefficients of
Eq.~(\ref{Aij}).  Introducing
\bea
D_0&=&{\cal F}_0,\qquad D_1={\cal F}_{12},\nn\\
D_2&=&{\cal F}_{13},\qquad D_3={\cal F}_{23},
\ena
the branch points are determined by
\bea
X_D=D_0^2-D_1^2-D_2^2-D_3^2,
\label{dice-X}
\ena
and
\bea
Y_D^2
&=&
(D_0-D_1-D_2-D_3)
\nn\\
&\times&
(D_0-D_1+D_2+D_3)
\nn\\
&\times&
(D_0+D_1-D_2+D_3)
\nn\\
&\times&
(D_0+D_1+D_2-D_3).
\label{dice-Y}
\ena
Equivalently,
\bea
X_D
=
{\cal F}_0^2-
\left({\cal F}_{12}^2+{\cal F}_{13}^2+{\cal F}_{23}^2\right),
\label{dice-XA}
\ena
and
\bea
Y_D^2
&=&
[{\cal F}_0-{\cal F}_{12}-{\cal F}_{13}-{\cal F}_{23}]
\nn\\
&\times&
[{\cal F}_0-{\cal F}_{12}+{\cal F}_{13}+{\cal F}_{23}]
\nn\\
&\times&
[{\cal F}_0+{\cal F}_{12}-{\cal F}_{13}+{\cal F}_{23}]
\nn\\
&\times&
[{\cal F}_0+{\cal F}_{12}+{\cal F}_{13}-{\cal F}_{23}].
\label{dice-YA}
\ena
The first factor in Eq.~(\ref{dice-YA}) is $\Delta_D$ and vanishes at
criticality. The elliptic parameter is
\bea
\kappa_D^2=\frac{X_D-Y_D}{X_D+Y_D}.
\label{dice-modulus}
\ena

For a six-coordinated spin the Gaussian reduction leaves the spin as a vertex of the effective triangular model. Hence no extra local normalization appears:
\bea
\lim_{R\to\infty}{\rm Pf}\,\Gamma_R^{(1)}
=
(1-\kappa_D^2)^{1/4},
\ena
and hence
\bea
M_1
=
(1-\kappa_D^2)^{1/8}
=
\left[\frac{2Y_D}{X_D+Y_D}\right]^{1/8}.
\label{dice-M-general}
\ena
Thus Eq.~(\ref{dice-M-general}) is the complete six-coordinated spin magnetization.

For a three-coordinated spin the odd source of Eq.~(\ref{local-star-source}) gives the normalized weight $\tanh(J_1\sigma_1+J_2\sigma_2+J_3\sigma_3)$. Define
\bea
\Theta_1&=&\tanh(J_1+J_2+J_3),\nn\\
\Theta_2&=&\tanh(J_1+J_2-J_3),\nn\\
\Theta_3&=&\tanh(J_1-J_2+J_3),\nn\\
\Theta_4&=&\tanh(-J_1+J_2+J_3).
\label{dice-theta}
\ena
Expanding it in the four odd spin monomials gives
\bea
\mathbb E[s_{2,3}\mid\sigma_1,\sigma_2,\sigma_3]
&=&c_1\sigma_1+c_2\sigma_2\nn\\
&&+c_3\sigma_3+c_{123}\sigma_1\sigma_2\sigma_3,
\label{dice-local-op}
\ena
with
\bea
c_1&=&\frac{\Theta_1+\Theta_2+\Theta_3-\Theta_4}{4},\nn\\
c_2&=&\frac{\Theta_1+\Theta_2-\Theta_3+\Theta_4}{4},\nn\\
c_3&=&\frac{\Theta_1-\Theta_2+\Theta_3+\Theta_4}{4},\nn\\
c_{123}&=&\frac{\Theta_1-\Theta_2-\Theta_3-\Theta_4}{4}.
\label{dice-c-coeff}
\ena
The two three-coordinated sublattices are inversion-related and have the same magnetization. Let
\bea
T_\triangle
=
\left\langle
\sigma_1\sigma_2\sigma_3
\right\rangle_{\triangle,+}
\label{dice-triplet}
\ena
be the elementary triplet of the reduced anisotropic
triangular model in the plus phase. Then
\bea
M_2=M_3
=
(c_1+c_2+c_3)\,M_1+c_{123}\,T_\triangle .
\label{dice-M23-general}
\ena
Equivalently,
\bea
M_2=M_3
=
A_D M_1+B_D T_\triangle ,
\ena
where
\bea
A_D&=&\frac{3\Theta_1+\Theta_2+\Theta_3+\Theta_4}{4},\nn\\
B_D&=&\frac{\Theta_1-\Theta_2-\Theta_3-\Theta_4}{4}.
\label{dice-AB}
\ena
Appendix~\ref{app:local-schur} represents $T_\triangle$ by a four-spin path Pfaffian. The same even Gaussian weight gives the triangular couplings from the four local 
weights
\bea
W_1&=&2\cosh(J_1+J_2+J_3),\nn\\
W_2&=&2\cosh(J_1+J_2-J_3),\nn\\
W_3&=&2\cosh(J_1-J_2+J_3),\nn\\
W_4&=&2\cosh(-J_1+J_2+J_3),
\ena
the even Gaussian boundary weight generates
\bea
e^{4K_{12}}&=&\frac{W_1W_2}{W_3W_4},\nn\\
e^{4K_{23}}&=&\frac{W_1W_4}{W_2W_3},\nn\\
e^{4K_{31}}&=&\frac{W_1W_3}{W_2W_4}.
\label{dice-schur-triangle}
\ena

Each triangular bond receives equal contributions from the two three-coordinated
sublattices, so $\widehat K_{ab}=2K_{ab}$, equivalently
\begin{equation}
\tanh\widehat K_{ij}=\frac{S_iS_j}{C_iC_j+C_k},
\qquad\{i,j,k\}=\{1,2,3\}.
\label{dice-triangular-couplings}
\end{equation}
With $H=C_1^2+C_2^2+C_3^2+2C_1C_2C_3-1$, the four coefficients in
Eq.~(\ref{Aij}) equal $2\mathcal N H$ times the standard triangular
spectral coefficients. The common scale cancels from Eq.~(\ref{dice-modulus}),
confirming the six-coordinated result.

A second star--triangle transformation closes the triplet. Set
\begin{align}
x_1&=\tanh\widehat K_{23},\quad
x_2=\tanh\widehat K_{31},\quad
x_3=\tanh\widehat K_{12},\nonumber\\
d_\triangle&=1+x_1x_2x_3,\qquad b_i=x_i+x_jx_k,\nonumber\\
\eta_i^2&=\frac{b_jb_k}{d_\triangle b_i},\qquad
\ell_i=\operatorname{artanh}\eta_i.
\label{dice-triplet-map}
\end{align}
Let $\mathsf A_\star,\mathsf B_\star$ denote Eq.~(\ref{dice-AB}) with
$J_i\to\ell_i$. The transformed honeycomb sublattices are inversion-related,
so the inserted-spin expectation gives
$M_1=\mathsf A_\star M_1+\mathsf B_\star T_\triangle$, hence
\begin{align}
T_\triangle&=\frac{1-\mathsf A_\star}{\mathsf B_\star}M_1,\nonumber\\
M_2=M_3&=\left[A_D+B_D\frac{1-\mathsf A_\star}{\mathsf B_\star}\right]M_1.
\label{dice-anisotropic-closed}
\end{align}
For positive finite couplings $\mathsf B_\star<0$; zero-coupling limits are
continuous. These equations determine all three ordered-phase sublattices,
with $M_D^{\rm av}=(M_1+2M_2)/3$.

For the dual kagom\'e couplings $e^{-2\bar J_i}=\tanh J_i$, direct substitution gives
\bea
{\cal F}_0&=&\Omega{\cal D}_0(\bar\lambda),\qquad
{\cal F}_{12}=\Omega{\cal D}_3(\bar\lambda),\nn\\
{\cal F}_{13}&=&\Omega{\cal D}_2(\bar\lambda),\qquad
{\cal F}_{23}=\Omega{\cal D}_1(\bar\lambda),
\ena
where
\bea
\Omega=\frac{64}{\displaystyle\prod_{i=1}^{3}
\update{(1+6 \bar\lambda_i^2+ \bar\lambda_i^4)}}.
\ena
Hence $X_D=\Omega^2X_K$ and $Y_D^2=\Omega^4Y_K^2$.
Kramers--Wannier duality maps the ordered dice phase onto the
disordered kagom\'e phase, so the physical square-root branches have
opposite signs, $Y_D=-\Omega^2Y_K$, and therefore
\bea
\kappa_D^2(J_1,J_2,J_3)
=
\frac{1}{\kappa_K^2(\bar J_1,\bar J_2,\bar J_3)}.
\ena
Thus
\bea
M_1^D(J_1,J_2,J_3)
=
\left[
1-\frac{1}{\kappa_K^2(\bar J_1,\bar J_2,\bar J_3)}
\right]^{1/8}.
\ena
This relates dice spin order to dual kagom\'e disorder order; it is not the physical kagom\'e spin magnetization.

For $J_i>0$ these formulas apply in the ordered region $\mathcal Q<0$ with $Y_D>0$ and continue to $M_a=1$ as $T\to0$; set $M_a=0$ for $\mathcal Q\geq0$. In duality formulas use the signed continuation $Y_D=-\operatorname{sgn}(\mathcal Q)\sqrt{Y_D^2}$. At criticality $Y_D=0$, $\kappa_D^2=1$, and all three magnetizations vanish with exponent $1/8$.

\subsection{Isotropic dice lattice.} Let us now put
\bea
J_1=J_2=J_3=J,
\;
C=\cosh(2J),
\;
S=\sinh(2J).\qquad
\ena
In this case
\bea
{\cal F}_{12}={\cal F}_{13}={\cal F}_{23}\equiv {\cal F},
\quad
{\cal F}=4\; {\cal N}\; S^2C(C+1),\qquad
\ena
and
\bea
{\cal F}_0
= {\cal N}
\Big[3+4C^3+(2C^2-1)^3 \Big].
\ena
Equations (\ref{dice-XA}) and (\ref{dice-YA}) then reduce to
\bea
X_D={\cal F}_0^2-3 {\cal F}^2,\;\;
%
%
Y_D^2
=
({\cal F}_0-3 {\cal F})({\cal F}_0+{\cal F})^3.\nn\quad
\ena
Using
\bea
{\cal F}_0-3 {\cal F}
=
2\mathcal N(C+1)^2
(2C^2-2C-1)^2,\nn\\
%
%
{\cal F}_0+{\cal F}
=
2\mathcal N(C+1)^2
(2C^2-2C+1)^2,\nn
\ena
one obtains in the ordered phase
%
\begin{align}
&Y_D=4\mathcal N^2(C+1)^4(2C^2-2C-1)(2C^2-2C+1)^3,\nonumber\\
&X_D+Y_D=128\mathcal N^2C^3(C-1)^3(C+1)^4(C^2-C+1).
\end{align}
%
Substitution into Eq.~(\ref{dice-M-general}) gives the compact result
\bea
	M_1(J)
	=
	\left[
	\frac{
		(2C^2-2C-1)
		(2C^2-2C+1)^3
	}{
		16C^3(C-1)^3(C^2-C+1)
	}
	\right]^{1/8},
\qquad
\label{dice-M-isotropic}
\ena
or, equivalently,
\bea
	M_1(J)
	=
	\left[
	1-
	\frac{(2C-1)^2}
	{16C^3(C-1)^3(C^2-C+1)}
	\right]^{1/8}. \qquad
\label{dice-M-isotropic-2}
\ena

For the isotropic three-coordinated sublattices, set
\bea
v=\tanh J .
\ena
The local reduction gives an isotropic triangular coupling
\bea
K_\triangle
=
\frac12\ln\left[
\frac{\cosh(3J)}{\cosh J}
\right]
=
\frac12\ln\left(
\frac{1+3v^2}{1-v^2}
\right).
\label{dice-Ktriangle-iso}
\ena
Accordingly,
\bea
M_1=M_\triangle(K_\triangle).
\ena
For the local three-coordinated spin one has
\bea
A_D
&=&
\frac34\left[\tanh(3J)+\tanh J\right],
\nn\\
B_D
&=&
\frac14\left[\tanh(3J)-3\tanh J\right].
\label{dice-AB-iso}
\ena
The triangular triplet is
\bea
T_\triangle=r_\triangle M_1,
\label{dice-triplet-ratio}
\ena
with exact ratio \cite{BaxterTriplet1975,BaxterChoy1989}
\bea
r_\triangle
&=&
\frac{3\eta^3-3\eta^2+3\eta-1}{2\eta^3},
\qquad \nn\\
\eta^2
&=&
\frac{e^{4K_\triangle}-1}{e^{4K_\triangle}+3}
=
\frac{2v^2(1+v^2)}{1+3v^4}.
\label{dice-rtriangle} 
\ena
Hence
\bea
M_2=M_3
=
\rho_3^D(J)\,M_1,
\label{dice-M23-isotropic}
\ena
with
\bea
\rho_3^D(J)
&=&
\frac34\left[\tanh(3J)+\tanh J\right]\nn\\
&+&
\frac14\left[\tanh(3J)-3\tanh J\right]
r_\triangle . 
\label{dice-rho3}
\ena
As $T\to0$, $r_\triangle,\rho_3^D\to1$, so all three dice magnetizations approach unity.
As a numerical check, Eqs.~(\ref{dice-M-isotropic}) and
(\ref{dice-rho3}) give
\bea
\begin{array}{c|ccc}
J
&
M_1
&
\rho_3^D
&
M_2=M_3
\\ \hline
0.55&0.984668&0.932698&0.918398\\
0.70&0.998375&0.971118&0.969540\\
0.90&0.999900&0.991065&0.990966
\end{array}
\label{dice-numerical-check}
\ena
These exact values show the stronger finite-temperature polarization of the six-coordinated sublattice.

Equation (\ref{dice-M-isotropic}) is valid for $J>J_c$, while
\bea
M_1(J)=0,\qquad J\leq J_c. \nn
\ena
The critical coupling obtained in Eq.(\ref{cc}),
\bea
C_c=\cosh(2J_c)=\frac{1+\sqrt{3}}{2},
\ena
satisfies
\bea
2C_c^2-2C_c-1=0,\nn
\ena
and therefore Eq.~(\ref{dice-M-isotropic}) gives
\bea
M_1(J_c)=0.\nn
\ena
In the opposite limit $J\rightarrow\infty$, one has
$C\rightarrow\infty$ and
\bea
M_1\rightarrow1, \nn
\ena
as required for the completely ordered ferromagnetic state.

It is useful to note that near the critical point
\bea
M_1\sim (J-J_c)^{1/8}, \nn
\ena
so that the usual two-dimensional Ising magnetization critical
exponent $\beta=1/8$ is reproduced.

\paragraph{kagom\'e--dice magnetization duality.}
For homogeneous dual couplings,
\bea
 e^{-2J_D}=\tanh J_K,
 \qquad \sinh(2J_K)\sinh(2J_D)=1,\qquad
 \label{kag-dice-duality-hom}
\ena
and the physical branches satisfy
\bea
 \kappa_D^2(J_D)=\frac{1}{\kappa_K^2(J_K)},\quad
 M_1^D(J_D)=\left[1-\frac{1}{\kappa_K^2(J_K)}\right]^{1/8}.\qquad
 \label{kag-dice-kappa-duality-hom}
\ena
This does not equate the physical spin magnetizations. Kramers--Wannier
duality maps spin order to disorder order,
\bea
 M_K(J_K)\longleftrightarrow\mu_D(J_D),\quad
 M_1^D(J_D)\longleftrightarrow\mu_K(J_K),\qquad
 \label{kag-dice-order-disorder}
\ena
whereas $M_K=\rho_K(1-\kappa_K^2)^{1/8}$. The dual coupling exchanges
ordered and disordered phases; the two three-coordinated dice spins also
require their own transformed operators.

 \section{SUMMARY}
  In this work we have developed a unified fermionic-field formulation
 of the two-dimensional Ising model on planar lattices with different
 local geometries. The construction starts from the high-temperature
 loop representation and the Kac--Ward prescription. Grassmann
 fermionic variables are associated with directed lattice links, while
 a change of the propagation direction is accompanied by the
 corresponding Kac-Ward rotation factor. In this way the geometrical
 properties of the lattice are incorporated directly into a quadratic
 fermionic action.
 
 After Fourier transformation, the statistical-mechanical problem is
 reduced to a finite-dimensional momentum-space fermionic matrix
 ${\cal A}({\bf k})$. Its determinant determines the excitation
 spectrum,
 \bea
 \det {\cal A}({\bf k})=0,\nn
 \ena
 while the appearance of a zero mode at ${\bf k}=0$ determines the
 critical coupling. Expansion of the same determinant near the zero
 mode gives the long-wavelength fermionic spectrum and the mass gap.
 Thus the same description gives the critical point and long-wavelength fermionic scales; exact spin decay lengths additionally require the full spectral polynomial and spin insertion.
 
 We first applied the construction to the square, honeycomb, and
 triangular lattices and reproduced their exact critical couplings.
 These examples demonstrate that the Kac--Ward phase associated with
 the rotation of fermionic trajectories provides a direct and
 systematic way of constructing the corresponding lattice fermion
 operators.
 
 A more nontrivial application was given for the anisotropic kagom\'e
 lattice. Its elementary cell contains three four-coordinated sites,
 which leads to a $12\times12$ fermionic matrix. For three independent
 couplings $J_1,J_2,J_3$ we obtained the complete critical surface
 \bea
 \sum_{i=1}^{3}\cosh(2J_i)
 -
 \prod_{i=1}^{3}\cosh(2J_i)
 -
 \prod_{i=1}^{3}\sinh(2J_i)
 =0. \qquad \; \nn
 \ena
 The corresponding spectral equation was derived for arbitrary
 anisotropy. Its expansion around zero energy and momentum shows
 explicitly that the fermionic mass vanishes on this surface, while
 the isotropic limit reproduces the known kagom\'e critical coupling.
 
The full kagom\'e spin Pfaffian contains the common spectral factor and the
local spin insertion. Appendix~\ref{app:local-schur} gives
\bea
\rho_i=\frac{2\sqrt{d\,a_1a_2a_3}}{d\,a_i+a_ja_k},\qquad
M_i=\rho_i(1-\kappa_K^2)^{1/8},\quad i=1,2,3,\nn
\ena
with $d=1+\lambda_1\lambda_2\lambda_3$ and
$a_i=\lambda_i+\lambda_j\lambda_k$. The star--triangle check yields
$\rho_i=\tanh(2L_i)$; isotropically this reproduces Naya's result and
$M=1-2u^4-12u^6+O(u^8)$.

 We have also constructed the fermionic action for the lattice dual to
 the kagom\'e lattice, the dice or $T_3$ lattice. Although its local
 geometry is substantially different, with one six-coordinated and two
 three-coordinated sites in the elementary cell, the model is again
 described by twelve directed fermionic degrees of freedom. The
 anisotropic critical condition takes the symmetric form
 \bea
 &&2-\cosh[2(J_1+J_2+J_3)]
 +\cosh[2(J_1+J_2-J_3)] \nn
 \\
 &&
 \hspace{2mm}
 +\cosh[2(J_1-J_2+J_3)]
 +\cosh[2(-J_1+J_2+J_3)]
 =0 ,\nn 
 \ena
 and is related to the kagom\'e critical surface by the usual Ising
 duality transformation.
 
 For the isotropic dice lattice the low-energy spectrum reduces to
 \bea
 k_0^2+k^2+m^2=0,\nn
 \ena
 with the fermionic mass vanishing at
 %
 $\cosh(2J_c)=\frac{1+\sqrt{3}}{2}$, as expected.
 %
 The Green-function analysis also gives the spontaneous magnetization
 of the six-coordinated sublattice in the form
 \bea
 M_1=(1-\kappa_D^2)^{1/8},\nn
 \ena
Here $\kappa_D^2=1$ at criticality and tends to zero as $T\to0$.
The same Pfaffian reduction gives
$M_2=M_3=A_D M_1+B_D T_\triangle$; Eq.~(\ref{dice-anisotropic-closed})
closes $T_\triangle$, and isotropically $M_2=M_3=\rho_3^D M_1$ with
$\rho_3^D\to1$ as $T\to0$.
 Under duality $M_1^D$ maps to kagom\'e disorder order, not to the physical kagom\'e spin magnetization $M_i=\rho_i(1-\kappa_K^2)^{1/8}$.  Both the
 kagom\'e and dice results reproduce the well-known two-dimensional Ising
 magnetization exponent
 \bea
 \beta=\frac18 \nn.
 \ena
 The main result of the present work is therefore not merely the
 determination of critical-temperature surfaces for particular fully anisotropic lattices. Critical temperatures of homogeneous Archimedean lattices have already been studied extensively in the works cited in the Introduction.
 Rather, the present formulation provides a common fermionic framework in which the
 partition function, excitation spectrum, critical surface, mass gap,
 fermionic Green functions, spin correlations, and spontaneous
 magnetization can all be obtained from the same lattice fermion
 operator.  The dependence on lattice geometry enters through the
 connectivity of directed links and the corresponding Kac--Ward
 rotation factors, while the subsequent calculations have the standard
 form of a free-fermion field theory. This makes the approach
 particularly suitable for systematic studies of anisotropic,
 inhomogeneous, frustrated and more general planar Ising lattices.
 {Moreover, the present formulation provides a natural starting point for studying Ising models on random lattices, an important problem in the physics of disorder.} 
 


\section*{Acknowledgements}
 The work  was  supported by the Armenian HESC grants  21AG-1C024 and  24FP-1F039.

\appendix

\section{Path-dependent spin Pfaffian and endpoint Schur complement}
\label{app:path-pfaffian}
For a finite planar graph let $x_e=\tanh J_e>0$. Choose directed edge
pairs $(e^+,e^-)$ and define $\mathsf K_s=I-\mathsf T_s$ by
\begin{equation}
(\mathsf T_s)_{ee'}=
\begin{cases}
\sqrt{x_ex_{e'}}e^{i\phi(e,e')/2},&\operatorname{head}(e)=\operatorname{tail}(e'),\ e'\ne\bar e,\\
0,&\text{otherwise}.
\end{cases}
\end{equation}
With edge reversal $\mathsf J$ and $\mathsf U_{ee}=e^{-i\theta_e/2}$,
\begin{equation}
\widehat K=i\mathsf U^\dagger\mathsf J\mathsf K_s\mathsf U
\label{eq:kw-real-skew}
\end{equation}
is real and antisymmetric. In canonical pair ordering,
$\operatorname{Pf}\widehat K=\mathcal P(x)$, with
$\mathcal P(x)=\sum_{\partial F=\varnothing}\prod_{e\in F}x_e$; the sign
is fixed by $\widehat K(0)=\bigoplus_eJ_2$.

Let a path $\gamma:a\to b$ contain edges $e_1,\ldots,e_m$, and let
$P_\gamma$ select both directed states of each path edge. Define
\begin{align}
Q_\gamma&=\bigoplus_{r=1}^m x_{e_r}^{-1}J_2,\nonumber\\
D_\gamma&=\bigoplus_{r=1}^m\sqrt{x_{e_r}^{-1}-x_{e_r}}\,I_2,\nonumber\\
\Gamma_\gamma&=Q_\gamma+D_\gamma P_\gamma^T\widehat K^{-1}P_\gamma D_\gamma.
\label{eq:explicit-spin-path-pfaffian}
\end{align}
Then
\begin{equation}
\boxed{\langle s_as_b\rangle=\operatorname{Pf}\Gamma_\gamma.}
\label{eq:physical-spin-pfaffian-result}
\end{equation}
The matrix dimension is $2m$; the contact blocks and path weights are
therefore part of the spin insertion and cannot be replaced by an endpoint
Green function.

Indeed, replacing $x_e\to x_e^{-1}$ on $\gamma$ gives a rank-$2m$
antisymmetric perturbation. The Pfaffian identity
\begin{equation}
\frac{\operatorname{Pf}(A+P\Delta P^T)}{\operatorname{Pf}A}
=\frac{\operatorname{Pf}(\Delta^{-1}+P^TA^{-1}P)}
{\operatorname{Pf}\Delta^{-1}}
\label{eq:pfaffian-low-rank-lemma}
\end{equation}
then yields Eq.~\eqref{eq:physical-spin-pfaffian-result}, including its
normalization. Equivalently,
\begin{equation}
\langle s_as_b\rangle=
\left\langle\prod_{e\in\gamma}
[x_e^{-1}-(x_e^{-1}-x_e)\chi_{e^+}\chi_{e^-}]
\right\rangle_{\widehat K}.
\label{eq:explicit-gaussian-spin-string}
\end{equation}

For the row-weighted matrix $\mathcal A=\mathsf T_r-I$ used in the main
text, $\mathsf W_{ee}=\sqrt{x_e}$ gives
\begin{equation}
\widehat K^{-1}=i\mathsf U^\dagger\mathsf W^{-1}
\mathcal A^{-1}\mathsf W\mathsf J\mathsf U.
\label{eq:green-to-real-skew}
\end{equation}
Thus the Bloch Green function supplies every entry of $\Gamma_\gamma$,
including the unit-cell translation carried by edge reversal.

Separating the endpoint pairs ($E$) from the interior pairs ($I$),
\begin{equation}
\Gamma_\gamma=\begin{pmatrix}\Gamma_{EE}&\Gamma_{EI}\\
-\Gamma_{EI}^T&\Gamma_{II}\end{pmatrix},\qquad
S_E=\Gamma_{EE}+\Gamma_{EI}\Gamma_{II}^{-1}\Gamma_{EI}^T,
\end{equation}
so
\begin{equation}
\boxed{\operatorname{Pf}\Gamma_\gamma=
\operatorname{Pf}\Gamma_{II}\operatorname{Pf}S_E.}
\label{eq:endpoint-schur-exact}
\end{equation}
For $m>2$, $\operatorname{Pf}\Gamma_{II}=\langle s_{v_1}s_{v_{m-1}}\rangle$.
Hence the path interior cannot be discarded; the local kagom\'e factor
comes from transforming the physical spin source together with the local
Gaussian block.

\section{Local Gaussian elimination with physical spin sources}
\label{app:local-schur}
For external Grassmann variables $\eta_i$, define
\begin{equation}
\mathcal W[f](\eta)=2^{-n}\sum_{s_i=\pm1}f(s_1,\ldots,s_n)
\prod_{i=1}^n(1+s_i\eta_i).
\label{eq:walsh}
\end{equation}
Multiplication by a physical spin is represented by
\begin{equation}
\mathcal W[s_if]=\mathsf S_i\mathcal W[f],\qquad
\mathsf S_i=\left[\prod_{j<i}(1-2\eta_j\partial_j)\right](\eta_i+\partial_i),
\label{eq:spin-op}
\end{equation}
where the parity string ensures $\mathsf S_i^2=1$ and mutual
commutativity.

For internal variables $\chi$ and
$\mathcal K=\left(\begin{smallmatrix}A&B\\-B^T&C\end{smallmatrix}\right)$,
Gaussian elimination gives
\begin{align}
\int D\chi\,e^{\mathcal S}
&=\operatorname{Pf}(A)e^{\eta^T(C+B^TA^{-1}B)\eta/2},\label{eq:schur}\\
\int D\chi\,(q^T\chi+\ell^T\eta)e^{\mathcal S}
&=[(\ell^T-q^TA^{-1}B)\eta]I_0(\eta).
\label{eq:odd-schur}
\end{align}
The second relation retains the odd physical-spin source.

\subsection{kagom\'e triangle, star, and decorating spin}
Set
\begin{align}
\lambda_i&=\tanh J_i,
\qquad d=1+\lambda_1\lambda_2\lambda_3,\nonumber\\
a_i&=\lambda_i+\lambda_j\lambda_k.
\end{align}
The normalized triangle boundary tensor is
\begin{equation}
\mathcal W[W_\triangle/\prod_i\cosh J_i]
=d+a_3\eta_1\eta_2+a_2\eta_1\eta_3+a_1\eta_2\eta_3.
\label{eq:triangle-boundary}
\end{equation}
A rescaling and permutation of the terminal Kac--Ward block gives
\begin{align}
A_\triangle&=\begin{pmatrix}
0&1&0&0&0&\lambda_2^{-1}\\
-1&0&\lambda_3^{-1}&0&0&0\\
0&-\lambda_3^{-1}&0&1&0&0\\
0&0&-1&0&\lambda_1^{-1}&0\\
0&0&0&-\lambda_1^{-1}&0&1\\
-\lambda_2^{-1}&0&0&0&-1&0
\end{pmatrix},\nonumber\\
B_\triangle&=i\begin{pmatrix}1&0&0\\1&0&0\\0&1&0\\0&1&0\\0&0&1\\0&0&1\end{pmatrix}.
\label{eq:Atriangle}
\end{align}
Direct calculation yields
\begin{align}
\operatorname{Pf}A_\triangle&=\frac{d}{\lambda_1\lambda_2\lambda_3},\nonumber\\
B_\triangle^TA_\triangle^{-1}B_\triangle
&=\frac1d\begin{pmatrix}0&a_3&a_2\\-a_3&0&a_1\\-a_2&-a_1&0\end{pmatrix}.
\label{local-triangle-schur}
\end{align}
This reproduces Eq.~\eqref{eq:triangle-boundary}; all congruences act on
the spin source as well, so normalized correlations are unchanged.

An equivalent three-leg star is represented by
\begin{align}
A_\star&=\begin{pmatrix}
0&1&1&0&1&0\\-1&0&0&0&0&0\\-1&0&0&1&1&0\\
0&0&-1&0&0&0\\-1&0&-1&0&0&1\\0&0&0&0&-1&0
\end{pmatrix},\nonumber\\
B_\star&=\begin{pmatrix}0&0&0\\t_1&0&0\\0&0&0\\0&t_2&0\\0&0&0\\0&0&t_3\end{pmatrix}.
\label{eq:Astar}
\end{align}
Its even and central-spin tensors are
\begin{align}
E_\star&=1+t_1t_2\eta_1\eta_2+t_1t_3\eta_1\eta_3+t_2t_3\eta_2\eta_3,\nonumber\\
O_\star&=t_1\eta_1+t_2\eta_2+t_3\eta_3+t_1t_2t_3\eta_1\eta_2\eta_3.
\label{local-star-source}
\end{align}
Matching the triangle kernel gives
\begin{equation}
t_jt_k=\frac{a_i}{d},\qquad
t_i^2=\frac{a_ja_k}{d\,a_i},\qquad t_i=\tanh L_i.
\end{equation}

After the adjacent triangles are transformed, the original kagom\'e spin
is a decorating spin between honeycomb spins $\tau_a,\tau_b$. With
$t=\tanh L_i$,
\begin{equation}
A_d=J_2,
\qquad B_d=\begin{pmatrix}-t&0\\0&t\end{pmatrix},
\qquad q^T=(1,1).
\label{eq:decor-block}
\end{equation}
Equations~\eqref{eq:schur}--\eqref{eq:odd-schur} give
\begin{align}
E_d&=1+t^2\eta_a\eta_b,
\qquad O_d=t(\eta_a+\eta_b),\nonumber\\
O_d&=\frac{\rho_i}{2}(\mathsf S_a+\mathsf S_b)E_d,
\qquad \rho_i=\frac{2t_i}{1+t_i^2}=\tanh(2L_i).
\label{local-decorating-insertion}
\end{align}
Thus $\mathbb E[s\mid\tau_a,\tau_b]=\rho_i(\tau_a+\tau_b)/2$, and for
two distinct kagom\'e spins on honeycomb edges $e=(a,b)$, $f=(c,d)$,
\begin{align}
\langle s_es_f\rangle_K
=\frac{\rho_i\rho_j}{4}
[\langle\tau_a\tau_c\rangle_h+\langle\tau_a\tau_d\rangle_h
+\langle\tau_b\tau_c\rangle_h+\langle\tau_b\tau_d\rangle_h].
\label{eq:global-spin}
\end{align}
This is Eq.~(\ref{kag-finite-Pf-reduction}); at large separation
$M_i=\rho_iM_h$.

\subsection{Dice three-coordinated spin}
For $v_i=\tanh J_i$, the same star block gives
\begin{align}
E_D&=1+v_1v_2\eta_1\eta_2+v_1v_3\eta_1\eta_3+v_2v_3\eta_2\eta_3,\nonumber\\
O_D&=v_1\eta_1+v_2\eta_2+v_3\eta_3+v_1v_2v_3\eta_1\eta_2\eta_3.
\label{eq:dice-tensors}
\end{align}
Write
\begin{equation}
O_D=(c_1\mathsf S_1+c_2\mathsf S_2+c_3\mathsf S_3
+c_{123}\mathsf S_1\mathsf S_2\mathsf S_3)E_D.
\label{eq:dice-operator}
\end{equation}
With $p_{ij}=v_iv_j$, coefficient matching gives
\begin{equation}
\begin{pmatrix}
1&p_{12}&p_{13}&p_{23}\\p_{12}&1&p_{23}&p_{13}\\
p_{13}&p_{23}&1&p_{12}\\p_{23}&p_{13}&p_{12}&1
\end{pmatrix}
\begin{pmatrix}c_1\\c_2\\c_3\\c_{123}\end{pmatrix}
=\begin{pmatrix}v_1\\v_2\\v_3\\v_1v_2v_3\end{pmatrix}.
\label{eq:c-system}
\end{equation}
Using Eq.~(\ref{dice-theta}),
\begin{align}
c_1&=(\Theta_1+\Theta_2+\Theta_3-\Theta_4)/4,\nonumber\\
c_2&=(\Theta_1+\Theta_2-\Theta_3+\Theta_4)/4,\nonumber\\
c_3&=(\Theta_1-\Theta_2+\Theta_3+\Theta_4)/4,\nonumber\\
c_{123}&=(\Theta_1-\Theta_2-\Theta_3-\Theta_4)/4.
\label{eq:c-solution}
\end{align}
Therefore
\begin{equation}
M_2=M_3=(c_1+c_2+c_3)M_1+c_{123}T_\triangle.
\end{equation}
The triplet can be written as an even path Pfaffian by adding a remote
fourth spin,
\begin{equation}
T_\triangle=\lim_{R\to\infty}
\frac{\operatorname{Pf}\Gamma^\triangle_{\gamma_{\{1,2,3,R\}}}}
{M_\triangle},
\label{local-triplet-even-Pf}
\end{equation}
and Eq.~(\ref{dice-anisotropic-closed}) evaluates this limit explicitly.

\section{Spectral moduli of the reduced spin models}
\label{app:spectral-link}
The local reductions fix the spin insertions; we now identify the Bloch
moduli with those of the reduced spin models.

\subsection{kagom\'e--honeycomb}
Let $v_i=t_i^2=\tanh K_i^h$. The normalized honeycomb coefficients are
\begin{align}
H_0&=1+v_1^2v_2^2+v_2^2v_3^2+v_3^2v_1^2,\nonumber\\
H_i&=2v_jv_k(1-v_i^2).
\end{align}
Direct substitution gives
\begin{equation}
\mathcal D_\mu=d^4H_\mu,
\qquad\mu=0,1,2,3.
\label{spectral-kag-honey-scale}
\end{equation}
With
\begin{align}
f_0&=1+v_1v_2+v_2v_3+v_3v_1,\nonumber\\
f_i&=1+v_jv_k-v_iv_j-v_iv_k,\nonumber\\
F&=\prod_{\mu=0}^3f_\mu,
\qquad B=16v_1^2v_2^2v_3^2,
\end{align}
one finds $X_h=B+F$ and $Y_h^2=(B-F)^2$. In the ordered region $B>F$,
\begin{equation}
\kappa_K^2=\frac{X_h-Y_h}{X_h+Y_h}=\frac{F}{B}.
\label{spectral-honey-modulus}
\end{equation}
This is the standard anisotropic honeycomb/triangular spin modulus; hence
$M_h=(1-\kappa_K^2)^{1/8}$.
\vspace{3mm}
\subsection{Dice--triangular}
Let $C_i=\cosh2J_i$, $S_i=\sinh2J_i$, and
$H=C_1^2+C_2^2+C_3^2+2C_1C_2C_3-1$. The reduced triangular couplings obey
$\tanh\widehat K_{ij}=S_iS_j/(C_iC_j+C_k)$, while

\begin{align}
\mathcal F_{ij}&=2\mathcal N H\sinh(2\widehat K_{ij}),\nonumber\\
\mathcal F_0&=2\mathcal N H\left[
\prod_{a<b}\cosh(2\widehat K_{ab})+
\prod_{a<b}\sinh(2\widehat K_{ab})\right].
\label{spectral-dice-triangle-scale}
\end{align}

Thus all four dice coefficients differ from the triangular coefficients by
the same scale $2\mathcal N H$, which cancels from $\kappa_D^2$. The same
modulus therefore gives the six-coordinated magnetization, while the odd
source above determines the other two sublattices.

\end{document}